\documentclass{iau-JDSS}

\usepackage{graphicx, natbib}
\usepackage{psfig}
\usepackage{epsfig}
\newcommand{\apjl}{ApJ}
\newcommand{\apj}{ApJ}
\newcommand{\nat}{Nature}
\newcommand{\aap}{A\&A}
\newcommand{\mnras}{MNRAS}
\newcommand{\msun}{\mbox{$M_{\odot}$}}
\newcommand{\Msun}{\mbox{$M_{\odot}$}}
\newcommand{\lsun}{\mbox{$L_{\odot}$}}
\newcommand{\Lsun}{\mbox{$L_{\odot}$}}
  
\newcommand{\rsun}{\mbox{$R_{\odot}$}}

\newcommand{\vinf}{\mbox{$v_{\infty}$}}
\newcommand{\vesc}{\mbox{$v_{\rm esc}$}}

\newcommand{\mdot}{\mbox{$\dot{M}$}}

\newcommand{\mtir}{\ensuremath{\dot{M}_{\rm tir}}}
\newcommand{\Gammae}{\ensuremath{\Gamma_{\rm e}}}

\title[Very Massive Stars]{Very Massive Stars in the local Universe}

\author[Vink et al.]   
{Jorick S. Vink$^1$,
Alexander Heger$^2$,
Mark R. Krumholz$^3$, 
Joachim Puls$^4$, 
S. Banerjee$^5$,
N. Castro$^6$,  
K.-J. Chen$^7$,
A.-N. Chen\`{e}$^{8,9}$,
P.A. Crowther$^{10}$, 
A. Daminelli$^{11}$, 
G. Gr\"afener$^1$,
J. H. Groh$^{12}$, 
W.-R. Hamann$^{13}$,
S. Heap$^{14}$,
A. Herrero$^{15}$,
L. Kaper$^{16}$,
F. Najarro$^{17}$,
L. M. Oskinova$^{13}$,
A. Roman-Lopes$^{18}$,
A. Rosen$^{3}$,
A. Sander$^{13}$,
M. Shirazi$^{19}$,
Y. Sugawara$^{20}$,
F. Tramper$^{16}$,
D. Vanbeveren$^{21}$, 
R. Voss$^{22}$,
A. Wofford$^{23}$,
Y. Zhang$^{24}$
\break \and the participants of JD2}

\affiliation{$^1$Armagh Observatory, College Hill, BT61 9DG, Armaghm Northern Ireland, 
UK \break email: jsv@arm.ac.uk\\
$^2$Monash Centre for Astrophysics School of Mathematical Sciences, Building 28, M401
Monash University, Vic 3800, Australia\\
$^3$Department of Astronomy \& Astrophysics, University of California, Santa Cruz, CA 95064, USA\\
$^4$Universit\"{a}ts-Sternwarte, Scheinerstrasse 1, 81679, Munchen, Germany\\
$^5$Argelander-Institut f\"ur Astronomie, Auf dem H\"ugel 71, D-53121, Bonn, Germany\\
$^6$Institute of Astronomy \& Astrophysics, National Observatory of Athens, I. 
Metaxa \& Vas. Pavlou St. P. Penteli, 15236 Athens, Greece\\
$^7$Minnesota Institute for Astrophysics, University of Minnesota, Minneapolis, MN 55455, USA\\
$^8$Departamento de F\'{\i}sica y Astronom\'{\i}a, Universidad 
de Valpara\'{\i}so, Av. Gran Breta\~na 1111, Playa Ancha, Casilla 5030, Chile\\
$^9$Departamento de Astronom\'{\i}a, Universidad de Concepci\'on, Casilla 160-C, Chile\\
$^{10}$Dept. of Physics \& Astronomy, Hounsfield Road, University of Sheffield, S3 7RH, UK\\
$^{11}$Departamento de Astronomia do IAG-USP, R. do Matao 1226, 05508-090 Sao Paulo, Brazil\\
$^{12}$Geneva Observatory, Geneva University, Chemin des Maillettes 51, CH-1290 Sauverny, Switzerland\\
$^{13}$Institute for Physics and Astronomy, University Potsdam, 14476 Potsdam, Germany\\
$^{14}$NASA Goddard Space Flight Center, Greenbelt, MD 20771, USA\\
$^{15}$Instituto de Astrofisica de Canarias and Universidad de La Laguna, E-38200 La Laguna, Spain\\
$^{16}$Astronomical Institute `Anton Pannekoek', University of Amsterdam, Science Park 904, PO Box 94249, 1090 GE Amsterdam, The Netherlands\\
$^{17}$Departamento de Astrofísica, Centro de Astrobiologia, (CSIC-INTA), Ctra. Torrejón a Ajalvir, km 4, 
28850 Torrejon de Ardoz, Madrid, Spain\\
$^{18}$Physics Department - Universidad de La Serena - Av. Cisternas,
1200 - La Serena - Chile\\
$^{19}$Leiden Observatory, Leiden University, P.O. Box 9513, 2300 RA Leiden, The Netherlands\\
$^{20}$Department of Physics, Faculty of Science \& Engineering, Chuo University, 1-13-27 Kasuga, Bunkyo, Tokyo 112-8551\\
$^{21}$Astrophysical Institute, Vrije Universiteit Brussel, Pleinlaan 2, 1050, Brussels, Belgium\\
$^{22}$Department of Astrophysics/IMAPP, Radboud University Nijmegen, PO Box
9010, NL-6500 GL Nijmegen, the Netherlands\\
$^{23}$Space Telescope Science Institute, 3700 San Martin Drive,
Baltimore, MD, 21218, USA\\
$^{24}$Department of Astronomy, University of Florida, Gainesville, FL 32611, USA}

\pubyear{2012}
\volume{Volume 16}  
\pagerange{119--126}
\date{?? and in revised form ??}
\jname{Highlights of Astronomy, Volume 16}
\editors{Thierry Montmerle, ed.}
\begin{document}

\maketitle

\begin{abstract}
Recent studies have claimed the existence of very massive stars (VMS) 
up to 300\,\msun\ in the local Universe. 
As this finding may represent a paradigm shift for the canonical 
stellar upper-mass limit of 150\,\msun, it is timely to discuss 
the status of the data, as well as the far-reaching implications of such objects. 
We held a Joint Discussion at the General Assembly in Beijing 
to discuss (i) the determination of the current masses of the most massive
stars, (ii) the formation of VMS, (iii) their mass loss, and (iv) their evolution 
and final fate. The prime aim was to reach broad consensus between observers and
theorists on how to identify and quantify the
dominant physical processes.
\end{abstract}

\firstsection % if your document starts with a section,
\section{Introduction}

The last decade has seen a growing interest in the study of the most 
massive stars, as their formation seems to be favourable 
in the early Universe at low metalliciy ($Z$), and is thought to involve 
a population of objects in the range 100-300\,\msun\ (Bromm et al. 1999; 
Abel et al. 2002). The first couple of 
stellar generations may be good candidates for the reionization of the Universe. 
Notwithstanding the role of the first stars, the interest in 
the current generation of massive stars has grown as well.
Massive stars are important drivers for the evolution 
of galaxies, as the prime contributors to the chemical and 
energy input into the interstellar medium (ISM) through  
stellar winds and supernovae (SNe).  
A number of exciting developments have 
recently taken place, such as the detection of a  
long-duration gamma-ray burst (GRB) at a redshift of 9.4, just 
a few hundred millions years after the Big Bang 
(Cucchiara et al. 2011). This provides convincing  
evidence that massive stars are able to form and die massive when 
the Universe was not yet enriched.  

The specific reason for holding this JD was the 
recent evidence for the existence and subsequent deaths of 
very massive stars (VMS) up to 300\,\msun. 
Gal-Yam et al. (2009) claimed the detection of a 
pair-instability SN (PSN) from a VMS. These explosions are 
thought to disrupt stars without 
leaving any remnants. 
Crowther et al. (2010) re-analyzed 
the most massive hydrogen-and nitrogen-rich Wolf-Rayet (WNh) 
stars in the center of R136, the
ionizing cluster of the Tarantula nebula in the Large Magellanic
Cloud (LMC). The conclusion from their analysis was that stars usually
assumed to be below the canonical stellar upper-mass limit of 
150\,\msun\ (of e.g. Figer 2005), 
were actually found to be much more luminous, and with initial 
masses up to $\sim$320\,\msun.
Prior to discussing the formation, evolution, and fate of VMS, and 
before we should explore the full implications of these findings, it is 
imperative to discuss the various lines of evidence for 
and against VMS.

VMS are usually found in and around young massive clusters, such as the 
Arches cluster in the Galactic centre and the local starburst 
region R136 in the LMC. Such clusters may harbor intermediate-mass
black holes (IMBHs) with masses in the range of several 
100 to several 1000\,\msun\ and may provide insight 
into the formation of supermassive black holes of order 
$10^5$\,\Msun. Young clusters are also relevant for 
the unsolved problem of massive star formation. 

For decades it was a struggle to form stars over 10\,\Msun, as radiation
pressure on dust grains might halt and reverse the accretion flow onto
the central object (e.g., Yorke \& Kruegel 1977).  Because of this
problem, astrophysicists have been creative in forming massive stars
via competitive accretion and merging in dense cluster environments
(e.g., Bonnell et al. 1998).  In more recent times several multi-D
simulations have shown that massive stars might form via disk
accretion after all (e.g., Krumholz et al. 2009; Kuiper et al.  2010).
In the light of recent claims for the existence of VMS 
in dense clusters, however, we should redress the issue of 
forming VMS in extreme environments.

Massive clusters may be so dense that
their early evolution is largely affected by stellar dynamics, and 
possibly form very massive objects via runaway collisions (e.g., Portegies Zwart et al.
1999; G\"urkan et al. 2004),
leading to the formation of VMS up to 1000\,\msun\ at the
cluster center, which may produce IMBHs at the end of their lives, but only 
if VMS mass loss is  
not too severe (see Belkus et al. 2007, Yungelson et al. 2008, Glebbeek et al. 2009, Pauldrach
et al. 2012).

VMS are thought to evolve almost chemically
homogeneously (Gr\"afener et al. 2011), implying that knowing 
the exact details of 
the mixing processes (e.g.,
rotation, magnetic fields) could be less relevant in comparison 
to their canonical 
$\sim$10-60\,\msun\ 
counterparts. Instead, the evolution and death of VMS is likely
dominated by mass loss. A crucial issue regards the relevance 
of episodes of super-Eddington, continuum-driven mass loss (such as occurs 
in Eta Carinae and other
Luminous Blue Variable (LBV) star eruptions), 
which might be able to remove large amounts of
mass -- even in the absence of substantial line-driven winds (see Sect.\,\ref{sec:massloss}).  

A final issue concerns the fate of VMS, and more specifically whether 
VMS end their 
lives as canonical Wolf-Rayet (WR) stars giving rise to Type Ibc SNe, or 
do they explode prematurely during the LBV phase? Might some of the most massive stars 
even produce PSNe? And how do PSNe compare to the general 
population of super-luminous SNe (SLSNe) that have recently been 
unveiled by Quimby et al. (2011), and are now seen out to high redshifts 
(Cooke et al. 2012).  
Such spectacular events can only be understood once we have obtained a basic knowledge 
of the physics of VMS. 

In Section \ref{sec:def} a definition for VMS is adopted, before we discuss
the evidence for and against super-canonical stars, i.e. objects above the 
traditional 150\,\msun\ stellar upper mass limit (Sect.\,\ref{sec:exist}).
After it is concluded that VMS probably exist, the next question to address is how to 
form such objects (Sect.\,\ref{sec:formation}). We then discuss the 
properties of VMS in Sect.\,\ref{sec:props}, before discussing their mass loss
(Sect.\,\ref{sec:massloss}), evolution and fate (Sect.\,\ref{sec:evol}). We end 
with the implications (Sect.\,\ref{sec:impl}) and final words. 

\section{Definition of VMS}
\label{sec:def}

Before we can discuss the evidences for and against VMS, one of 
the very first issues we discussed during the JD was what actually constitutes 
a ``very'' massive star. One may approach this in several different ways. 
Theoretically, ``normal'' massive stars with masses above $\sim$8\,\msun\ are those that 
produce core-collapse SNe (Smartt et al. 2009), but what happens at the upper-mass end 
(Nomoto 2012)? 

Above a certain critical mass, one would expect the occurrence of PSNe, and ideally 
this could be the lower-mass limit for the definition of our VMS. 
However, in practice this number is not known a priori due to mass loss, and as a result 
the initial and final masses are not the same. In other words, the initial main-sequence 
mass for PSN formation is model-dependent, and thus somewhat arbitrary. 
Moreover, there is the complicating issue of pulsational pair-instability 
supernova (Puls-PSN) at masses below those of full-fedged PSNe 
(e.g. Woosley et al. 2007). One could of course 
resort to the mass of the helium (He) core for which objects 
reach the conditions of electron/positron pair-formation instability.
Heger showed this minimum mass to be $\sim$40\,\Msun\ to encounter 
Puls-PSN and $\sim$65\,\Msun\ to encounter PSNe (see also Chatzopoulus \& Wheeler 2012). 

An alternative definition could involve the spectroscopic 
transition between normal main-sequence O-type 
stars and hydrogen-rich Wolf-Rayet stars (of WNh type), which have also been shown 
to be core H burning main-sequence objects. 
However, such a definition would also be model dependent. 

For these reasons, we took the decision to follow a more pragmatic approach: 
we consider stars to be {\it very} massive when the initial masses are $\sim$100\,\Msun, or higher. 

\section{On the existence of VMS}
\label{sec:exist}

With this definition, the question of whether {\it very} massive stars exist can 
convincingly be addressed with the one answer: {\it yes} they do! 
However, the pertinent issue for our JD was whether the widely 
held ``canonical'' upper-mass limit of 150\,\Msun\ has recently been superseded.  

Paul Crowther gave the first invited review presenting evidence for initial masses as high as 
320\,\msun. Crowther provided historical context to some 
of the astronomical community's sceptism regarding such high masses in R136, in particular 
the spectacular claim for the existence of a 2500\,\Msun\ star R136 in the 30 Doradus region 
of the LMC some 3 decades ago (e.g. Cassinelli et al. 1981). 
Higher spatial resolution showed that R136 was actually not a single supermassive star, but 
the object broke up into a young cluster containing significantly lower mass objects, including 
the current record holder R136a1 (as well as R136a2, R136a3, and R136c).
% fig1
\begin{figure}
\begin{center}
\includegraphics[width=1.\columnwidth,clip]{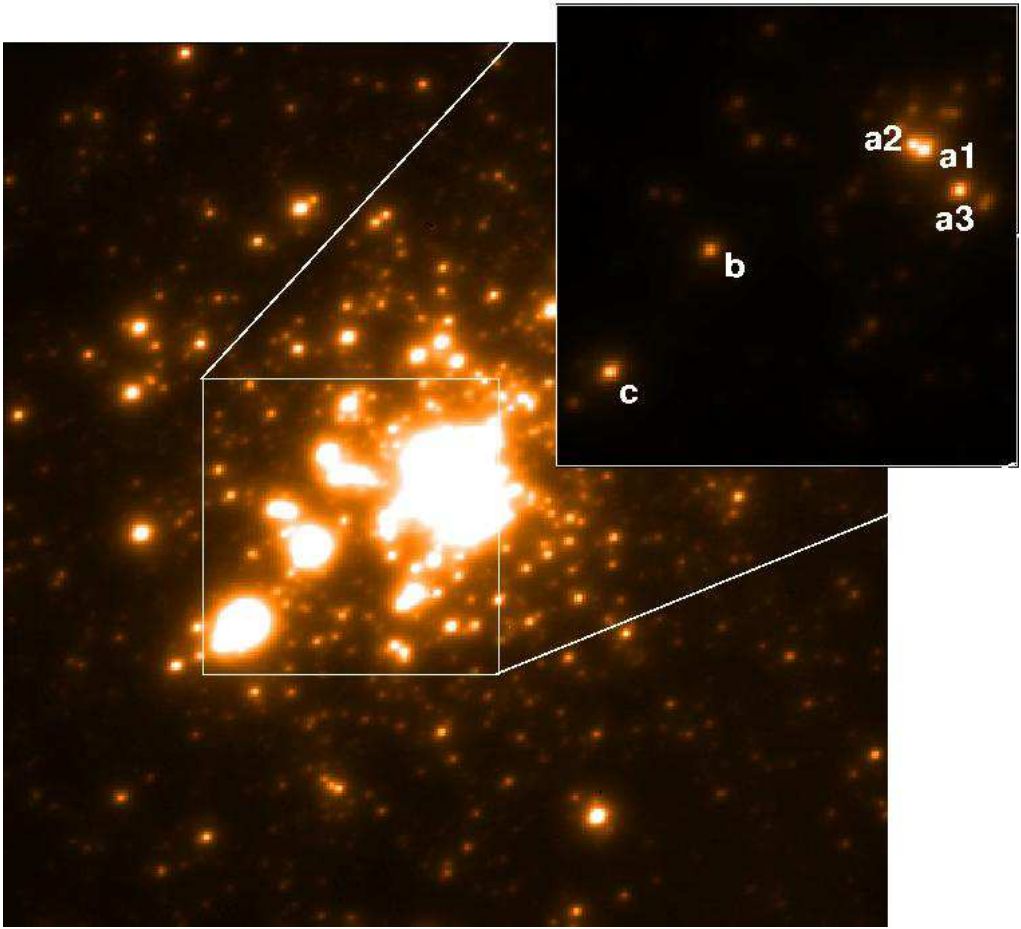}
\caption{
VLT MAD K$_{s}$-band 12 $\times$ 12 arcsec (3 $\times$ 3 parsec for the 
LMC distance of 49 kpc) image of R136 (Campbell et al. (2010) in conjunction 
with a view of the central 4 $\times$ 4 arcsec (1 $\times$ 1 parsec) in which the 
very massive WN5h stars discussed are labelled (component b 
is a lower mass WN9h star). Relative photometry agrees closely with 
integral field SINFONI observations of Schnurr et al. 2009). 
See Crowther et al. (2010) for details.}
\label{fig:MAD}
\end{center}
\end{figure}
Therefore, over the last few decades, there was a consensus of a 150\,\msun\ stellar upper mass limit 
(Weidner \& Kroupa 2004; Figer 2005; Oey \& Clarke 2005, Koen 2006), 
although the accuracy of this magic number of 150 was low (Massey 2011). 
 
Crowther et al. (2010) re-analyzed the photometric and spectroscopic data of the VMS 
in R136. In comparison to the older WFPC2 data, they used 
ground-based adaptive optics photometry (see Fig.\,\ref{fig:MAD}). 
In combination with their spectral analysis using the {\sc cmfgen} non-LTE 
atmosphere code of 
Hillier \& Miller (1998), this lead to higher estimates for 
effective temperatures and bolometric corrections. In conclusion they 
claimed that the R136 cluster hosts several stars with masses as high 
as 200-300\,\msun. 

Crowther et al. also performed a ``sanity check'' on similar WNh objects 
in the Galactic starburst cluster NGC\,3601. 
Although these objects were fainter, and less massive, than
those in R136, the advantage was the available dynamical mass estimate by 
Schnurr et al. (2008) of the binary object NGC\,3601-A1 of 116 $\pm$ 31 $+$ 89 $\pm$ 16 \Msun.
This is important as the least model-dependent way to obtain stellar masses is 
through the analysis of the light-curves and radial velocities 
induced by binary motions. We also note that Rauw et al. (2004) and Bonanos 
et al. (2004) found both components of the eclipsing Wolf-Rayet binary WR20a 
to be particularly massive, with 83 $\pm$ 5 and 82 $\pm$ 5\,\Msun, with small 
error bars. 

During the lively discussion that followed Paul's review, some attendants argued that 
the luminosities derived by Crowther et al. are uncertain
and these ``single'' objects might actually contain multiple sources. Whilst 
short-period binaries were not detected by Schnurr et al. (2009), longer period binaries 
are harder to exclude. One of the additional arguments by Crowther et al. was that X-rays
have not been detected, whilst they may have been expected on the basis of colliding wind binary (CWB) 
simulations by Pittard \& Stevens (2002). 
Oskinova however countered this argument on the basis that empirically 
a low X-ray luminosity cannot serve as a robust argument against a binary nature (see Sect.\,\ref{sec:massloss}). 

\begin{figure}
\includegraphics[scale=0.35]{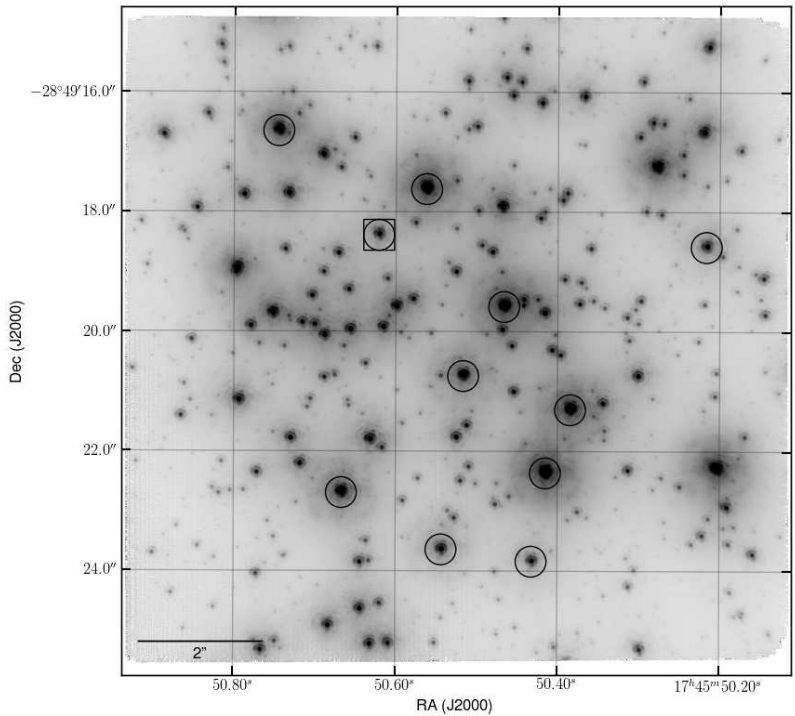}\hspace{0.2in}
\caption{
\label{arches}
K-band mosaic of the core field of the Galactic Arches cluster at a resolution of 
50 mas and with a sensitivity of $K_{\rm lim} = 20.6$mag. The figure is taken 
from Clarkson et al. (2012).
\label{fig:arches}}
\end{figure}

Najarro noted that even the best image so far of the Arches cluster with Keck (Fig.\,\ref{fig:arches}). 
has a limited spatial resolution of 50 milli-arcsec (mas), which corresponds to roughly 1/10 of the
diameter of the circles marking the PSF reference stars in the figure. 
Given that the LMC is almost 7 times further away, 
and that the VLT is smaller than Keck, the circles from Fig.\,\ref{fig:arches} 
would roughly correspond to the spatial
resolution achieved with the VLT if the Arches clusters was in the LMC. 
In other words, the Arches stars would effectively ``merge'' with surrounding objects. 
This analogy suggests that we cannot exclude the possibility 
that the bright WNh stars in the R136 cluster 
core could still break up into several lower-mass WNh stars. 
However, for example a $\sim$300\Msun\ star could at best break up into a pair of $\sim$150\msun\ 
stars (given the shallow slope of the stellar luminosity to mass ratio at the high-mass end). 

Moreover, Bestenlehner noted that 
there is a near-identical twin of R136a3 WNh star in 30 Doradus: VFTS\,682. 
Its key relevance is that it is found 
in isolation from the R136 cluster, some 30 pc away. For this object 
line-of-sight 
contamination is far less likely than for the R136 core stars. VFTS\,682 
thus offers another sanity check on the reliability of the 
luminosities for the R136 core stars. 
Bestenlehner et al. (2011) 
derived a high luminosity of log($L/\lsun$) $= 6.5$ and a 
present-day mass of 150\,\msun\ for VFTS\,682. This implied 
an initial mass on the zero-age main sequence (ZAMS) higher than the 
canonical upper-mass limit -- likely $\sim$ 200\,\Msun.  

In his talk Hamann showed results from another extremely luminous WN star 
in the Galactic center region: the Peony star (Barniske et al. 2008). 
The luminostiy of this star is determined from spectral analysis as
$\log(L/L_\odot)=6.5\pm 0.2$ and initial mass between
$150 \leq M_{\rm u}/M_\odot \leq 200$. The star is located above the Humphreys-Davidson 
limit, in the region populated by the LBV stars.
However, the hydrogen content is lower in WR\,102ka compared to the Pistol star, while 
helium is higher. This indicates a more advanced evolutionary
stage of the former, compared to ``normal'' LBV stars.

In summary, whilst one cannot exclude that the object R136a1 claimed to be $\sim$300\,\msun\ in 
the R136 cluster might still turn out to ``dissolve'' when higher spatial resolution observations were to 
become available, a number of sanity checks involving binary dynamics and ``isolated'' objects make it 
rather convincing that stars with ZAMS masses of 200\,\msun\ exist. 
In any case, no fundamental reason was identified why ``150'' would be a magic number. 
 
\section{Formation of VMS}
\label{sec:formation}

The key question to address was what is so special about the formation of 
250\,$\Msun$ stars in comparison to 'normal' $50\,\msun$ O-type stars? 
Can these VMS only form inside dense cluster environments?
Other relevant questions involve the time-scale and evolutionary 
stages when they finish formation/accretion. How fast do they rotate?
What VMS fraction is in binaries, and what are the binary properties?
Are the most massive stars all the result of stellar mergers?

\subsection{Theoretical considerations}

According to our second invited review speaker, Mark Krumholz, VMS formation 
is not a fundamentally different problem from the 
formation of massive stars in general. From an observational standpoint VMS 
in clusters appear as part of a 
continuous initial mass function (IMF), with no special features that mark them off as different from the 
remainder of the stellar population. From a theoretical standpoint, both the
 radiative and wind luminosities of stars are increasing functions of mass, but 
from the standpoint of star formation there is no natural dividing line that would 
put a star of, for example, 50 $M_\odot$ into one category and a star of 
250 $M_\odot$ into another. For this reason, it makes sense \textit{not} to treat 
the formation of very massive stars as a separate problem, and instead to embed 
it in the broader context of forming the upper end of the IMF.

\begin{figure}
\includegraphics{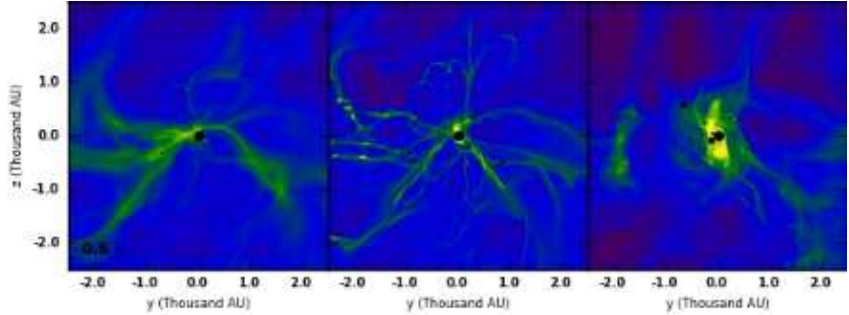}
\caption{
\label{krumholz_fig1}
Column density projections from three simulations of a collapsing high 
mass core, from \citet{myers12a}. All three simulations use identical 
initial conditions, but the leftmost one includes radiation and magnetic 
fields, the center one includes magnetic fields but not radiative transfer, 
and the rightmost one includes radiation but not magnetic fields. Note the 
dramatic reduction in number of stars (black circles) in the radiation plus 
magnetic fields case. The color scale runs from $10^{-0.75} - 10^{3.25}$ g cm$^{-2}$.
}
\end{figure}

There are three main challenges to forming massive stars: fragmentation, binarity, 
and radiation pressure. The fragmentation problem is simple to state: we certainly 
see gas over-densities (``cores") with masses of $\sim 100$ $M_\odot$ or larger and 
radii of $\sim 0.1$ pc or less that seem promising sites for massive star 
formation \cite[e.g.][]{beuther05b,bontemps10a}. However, given these masses 
and radii, and typical molecular cloud temperatures of $\sim 10$ K, the Jeans 
mass is below $1$ $M_\odot$, so how can the objects avoid fragmenting and collapse 
to form an object containing 100 Jeans masses or more? The answer seems to be that 
the classical Jeans analysis of an isothermal gas does a rather poor job at
 predicting the behavior of a radiatively heated, magnetized fluid. Once small 
stars form and begin to accrete, their radiation heats the gas around them and 
suppresses fragmentation in it \citep{krumholz06b, krumholz07a, krumholz08a}. 
Magnetic fields also make it much more difficult for the gas to fragment 
\citep{hennebelle11a}, and the combined effects of radiation and magnetic 
fields seems to be particularly effective, as seen in Fig.\,\ref{krumholz_fig1} 
\citep{commercon11a, myers12a}. 
These effects together seem to resolve the 
fragmentation problem, indicating that massive stars can form from the 
direct collapse of massive protostellar cores with properties similar 
to those observed.

The second challenge is explaining why so many massive stars appear to be 
binaries (e.g. Sana et al. 2012). 
The observed multiplicity fraction rises very sharply with stellar mass, 
reaching near unity for O and B stars (excluding runaways -- e.g.~\citealt{brown01a}). 
Radiation-hydrodynamic 
simulations of star cluster formation appear able to replicate the observed 
dependence of multiplicity on primary mass \citep{bate12a, krumholz12b}, and 
digging into the physical origin for this result indicates that it combines 
two effects. The first is simple N-body processing: close encounters 
between stars in young clusters tend to put the most massive members into 
binaries even if they are not born that way, while stripping companions 
from less massive stars. The second is disk fragmentation, as explored by 
\citet{kratter06a} and \citet{kratter08a, kratter10a}. Massive stars form 
with high accretion rates, and these accretion rates tend to produce disks 
with masses that can approach that of the primary. When this happens, 
disks are likely to fragment, and a common outcome of this process is 
that the disk produces a massive companion to the primary.

\begin{figure}
\includegraphics[scale=0.24]{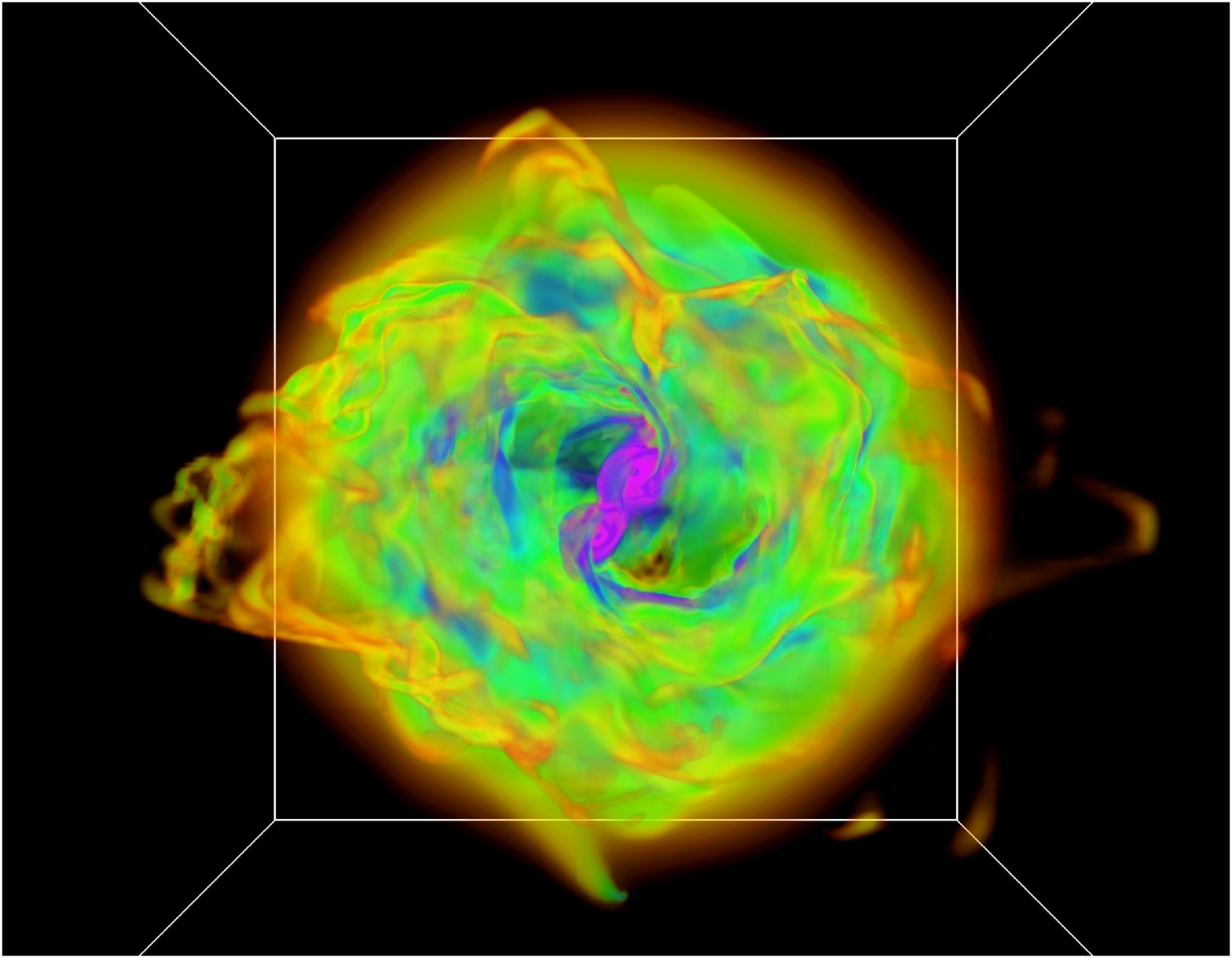}\hspace{0.2in}
\includegraphics[scale=0.24]{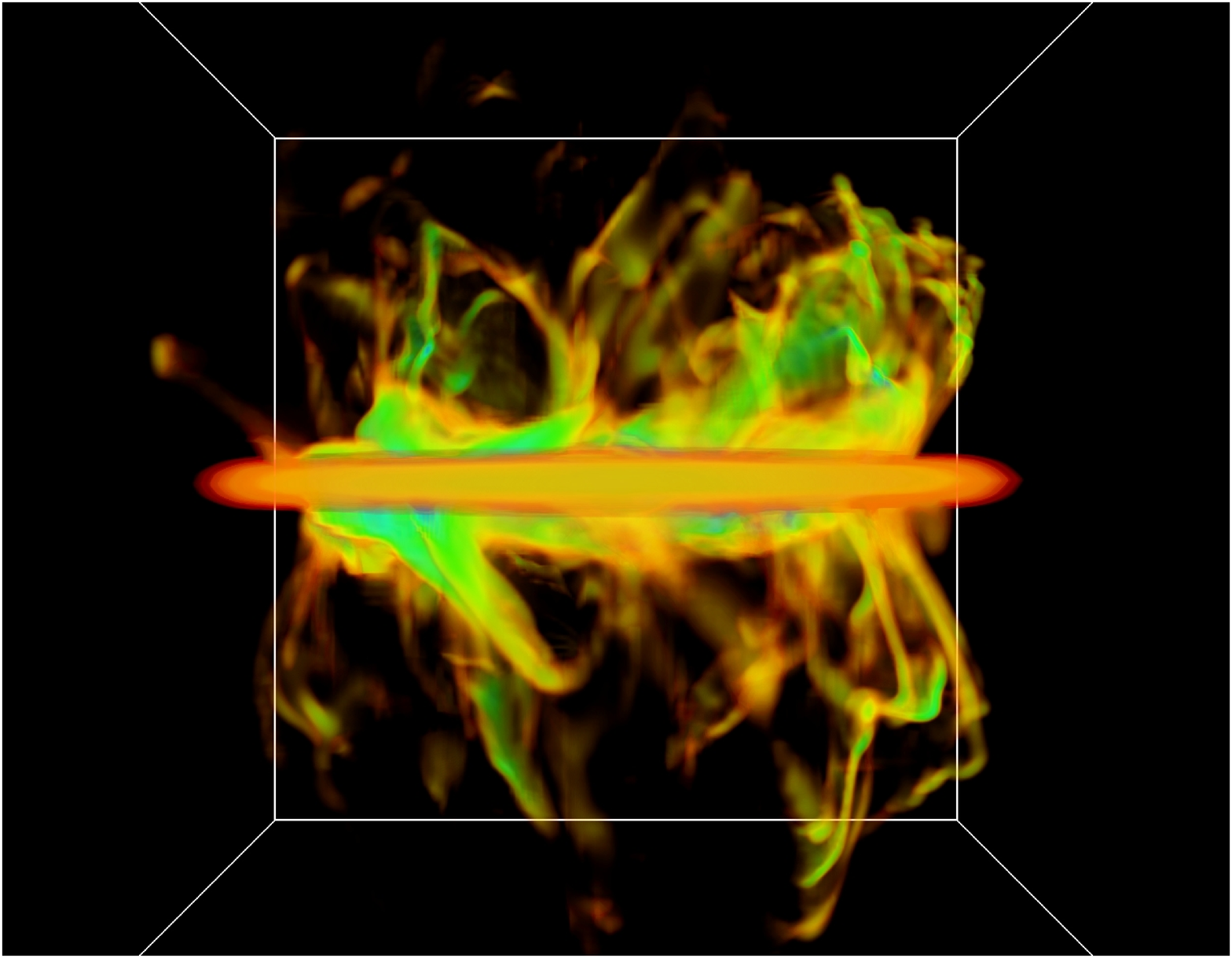}
\caption{
\label{krumholz_fig2}
Volume renderings of a simulation from \citet{krumholz09c} 
involving the density field in a $(4000\mbox{AU})^3$ region at 55.0 kyr of evolution. 
The color scale is logarithmic and runs from $10^{-16.5}-10^{-14}$ g cm$^{-3}$. 
The left panel shows the polar view, and the right one denotes the edge-on view. 
The figure highlights how 
Rayleigh-Taylor instability fingers channel matter onto a massive binary star system.
}
\end{figure}

The third challenge is radiation pressure. Dusty interstellar gas has a 
high opacity, and the radiation force is proportional to this opacity. As a 
result, the radiative force exerted by a star's light can exceed its 
gravitational force for all stars larger than $\sim 20$ $M_\odot$ 
\citep{wolfire87a, krumholz09c}. However, recent numerical and analytic 
work has shown that this problem is mostly an illusion; the radiation 
pressure barrier can be circumvented in numerous ways. First, in the 
presence of an accretion disk, radiation can be beamed away from the 
bulk of the accreting matter \citep{nakano95a, jijina96a, yorke02a,kuiper10b, kuiper11a}. 
Second, radiation-driven Rayleigh-Taylor instabilities can break up bubbles 
of radiation and allow matter to accrete through optically-thick fingers 
\citep{krumholz09c, jacquet11a}. Figure \ref{krumholz_fig2} shows an 
example. Third, protostellar jets can punch holes in accreting cores 
that allow radiation to leak out, reducing the net radiation force in
 most directions and allowing accretion to continue even though, averaged 
over $4\pi$ sr, the radiation force is stronger than gravity 
\citep{krumholz05a, cunningham11a}. The takeaway message from all of these 
simulations is that radiation pressure poses no barrier to the formation 
of stars up to arbitrary masses.

Given that modern theoretical models have removed all of the serious 
objections to forming massive stars in the same way that low mass stars 
form, i.e.~by accretion from a collapsing gas core, there is no need to 
resort to exotic processes like stellar collisions \citep[e.g.][]{bonnell98a}. 
This does not mean, however, that stellar collisions cannot happen and 
also contribute to the massive star population. The question is: under 
what circumstances do we expect collisions to be important? The most 
recent and comprehensive papers to address this question are
 \citet{moeckel11a} and \citet{baumgardt11a} who both conducted N-body 
simulations of stars confined by a gaseous potential, and who came to 
similar conclusions. They find that the collisional formation of very massive stars 
is only significant if the stellar density is extremely high, in excess 
of $10^7$ pc$^{-3}$, with surface densities reaching $10^5$ pc$^{-2}$. These 
numbers are so high that even the Arches cluster would not be expected to 
have significant contributions to its massive star population by collisions. 
Moreover, if collisions are significant, they do not produce an IMF that 
looks like what we observe. Instead, because collisions tend to occur 
among the most massive stars, they produce an IMF that has a Salpeter-like 
slope at intermediate masses, then a deficit of stars at higher masses, 
and finally one or a few very massive stars. IMFs with dips of this sort
 have not been observed, again suggesting that collisions are not likely 
to be significant contributors to the massive star populations.

\subsection{The potential role of dynamically induced mergers}

An alternative VMS formation scenario 
was presented by Sambaran Banerjee (with Pavel Kroupa and Seungkyung
Oh; Banerjee et al. 2012a) who argued that super-canonical 
stars can be formed out of a dense stellar population -- 
with a canonical IMF {\it and} with a 150\,\msun\ upper limit --
through dynamically induced mergers of the most massive binaries. Banerjee et al. performed 
direct N-body computations (NBODY6; Aarseth 2003) of a fully mass-segregated
star cluster mimicking R136 in which all the massive stars
are in primordial binaries. Banerjee et al. account for the mass evolution of 
the super-canonical stars and the resulting shortened ($\approx 1.5$ Myr)
lifetimes in their super-canonical phases using stellar evolutionary models 
by K\"ohler \& Langer (2012) that incorporate Vink et al. (2000) mass-loss rates for the main sequence 
and Hamann et al. (1995) for the He burning WR phase.

Banerjee et al. find that super-canonical stars begin to
form via dynamical mergers of massive binaries from $\approx 1$ Myr cluster age, obtaining 
stars with initial masses up to $\approx250\,\msun$. Multiple super-canonical stars are found to remain
bound to the cluster simultaneously within a super-canonical lifetime.
Banerjee also noted that some of these objects can be formed at runaway velocities
which escape the cluster at birth. For instance, the most massive apparently isolated 
WNh star VFTS\,682 might be an expected slow runaway (Banerjee et al. 2012b; see also 
Fujii \& Portegies-Zwart 2012).  

The Banarjee et al. models indicate that had super-canonical stars formed primordially alongside the rest of the R136 cluster,
i.e. violating the canonical upper limit, they would have evolved below the canonical $150\,\msun$ limit by
$\approx 3$ Myr, the likely age of the bulk of R136 according to Andersen et al. (2009). In other words, 
Banerjee et al. argue that primordially-formed super-canonical stars should not be observed
at the present time in R136, whilst it is quite plausible that a collection of dynamically 
formed super-canonical VMS would be observed in the centres of young massive starburst clusters.

A fully self-consistent N-body computation incorporating detailed 
accurate evolutionary and mass-loss recipes would be needed to confirm these scenarios.

\subsection{Rotation rates as a constraint on massive star formation}

Returning to the more conventional ways of forming massive stars, 
Anna Rosen addressed the question of what sets the initial rotation
rates of massive stars.
The physical mechanisms that set the initial massive star
rotation rates are a crucial unknown in current star-formation theory.
Observations of young, massive stars provide evidence that they form in
a similar fashion to their lower mass counterparts.
The magnetic coupling between a star and its accretion disk
may be sufficient to spin down low-mass pre-main-sequence (PMS) stars to
well below breakup at the end stage of their formation when the accretion rate
is low. However, Anna showed that these magnetic torques are insufficient to spin
down massive PMS stars due to their short formation times and high accretion
rates. Anna developed a model for the angular momentum evolution of stars over a
wide range in mass, considering both magnetic and gravitational torques. She
finds that magnetic torques are unable to spin down either low-mass or high-mass
stars during the main accretion phase, and that massive stars cannot be
spun down significantly by magnetic torques during the end stage of their
formation either. Spin-down occurs only if massive stars' disk lifetimes are
substantially longer or their magnetic fields are much stronger than current
observations suggest (Rosen et al. 2012).

\subsection{Observations of massive star formation}

Heavy extinction hides the birthplaces of massive stars from view, and the short 
formation timescales set a strong limitation to the sample of objects that can be studied. 
So far, our physical knowledge of massive young stellar objects (YSOs) 
has been derived from near-IR imaging and spectroscopy, revealing 
populations of young OB-type stars, some still surrounded by a disk, others apparently 'normal' main 
sequence stars powering H~{\sc ii} regions. The most important spectral features of
OB-type stars are however located in the ultraviolet (UV) and optical range. 

\begin{figure}
\includegraphics[width=1.\columnwidth,clip]{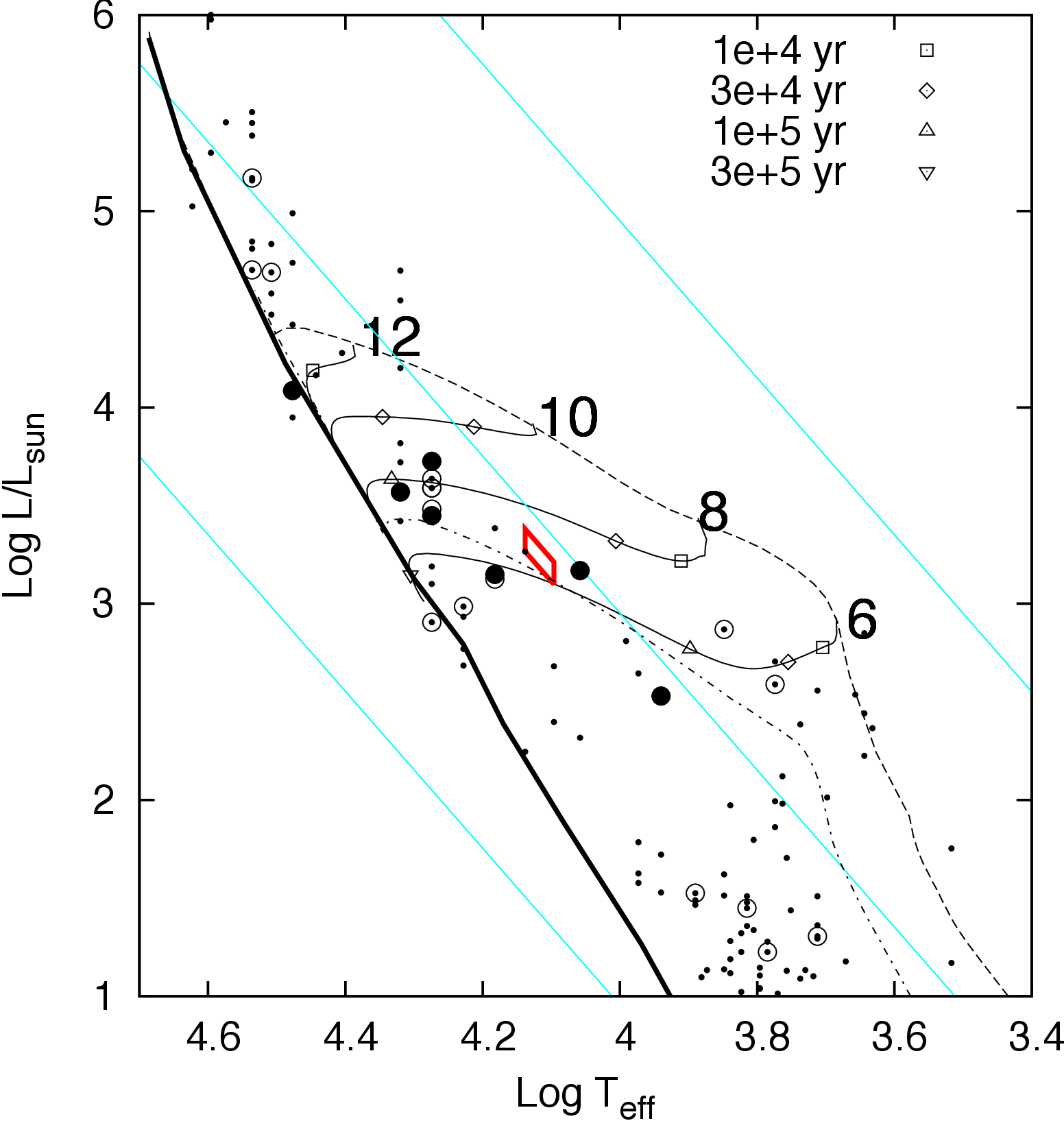}
\caption{
\label{ochsendorf}
The location of B275 (red parallelogram) in the HRD next to PMS tracks from Hosokawa et al. (2010) with the ZAMS mass
labeled and open symbols indicating lifetimes. The thin dashed and thin dot-dashed lines are the birth lines for accretion
rates of $10^{-4}$~M$_{\odot}$~yr$^{-1}$ and $10^{-5}$~M$_{\odot}$~yr$^{-1}$, respectively; the thick solid line is the ZAMS
(Schaller et al. 1992). The filled and open circles represent stars in M 17 for which a spectral type has been determined
(Hoffmeister et al. 2008); dots are other stars in M 17. B275 is on
its way to becoming a 6 − 8\,\Msun\ ZAMS star, so far one of the most massive pre-main-sequence star known. The 
figure has been adapted from Ochsendorf et al. (2011).
}
\end{figure}

Kaper (with Ellerbroek, Ochsendorf and Bik) showed that with the new optical/near-infrared spectrograph X-shooter on ESO's 
Very Large Telescope (VLT), it is possible to extend the spectral coverage of these massive YSOs into the optical range. 
First results are very promising, although they seem to probe the intermediate-mass ($\sim 2 - 8$~M$_{\odot}$) range 
rather than the massive star range. Ellerbroek et
al. (2011) discovered a jet (HH 1042) produced by the massive YSO nr292 in the massive star 
forming region RCW\,36, demonstrating that the object is
still actively accreting. The mass of this star is likely less than 6~M$_{\odot}$ and remains uncertain as photospheric
features are not detected. The first firm spectral classification of B275, a massive YSO in M17, results in its precise location on a PMS 
track for a $\sim7$\,\msun\ star (see Fig.\,\ref{ochsendorf}; Ochsendorf et al. 2011), has the size of a bloated giant, 
as predicted by models of Hosokawa et al. (2010), and is still surrounded by a disk. 

It remains unclear whether the progenitors of the most massive VMS will be detectable this way.

\subsubsection{Radiation Transfer Modeling: From massive to very massive star formation}

In order to properly interpret the observations of massive protostars, 
Zhang et al. (with Tan, McKee, and de Buizer) presented a radiation transfer (RT)
model for a massive core in a high pressure environment that forms a 
massive star through core accretion (Zhang \& Tan 2011; Zhang, Tan \& McKee, 2012, ApJ submitted). 
This RT model is 
based on the Turbulent Core model of McKee \& Tan (2003): massive stars form in dense clumps with 
high surface density. Assuming the rotational-to-gravitational
energy ratio is 2\% in the core, whilst the disk has a diameter of $\sim$1000 AU. 
The high accretion rate leads to a disk mass comparable to the stellar mass. 
In Zhang et al. (2012), the treatment of the disk was improved by allowing radially varying accretion rates 
due to a supply of mass and angular momentum from the
infall envelope and their loss to the disk wind. 
The transfer of accretion power to mechanical power of the wind was also accounted for. 
An approximate disk wind solution was developed partly based on the Blandford \& 
Payne (1982) model. The simulation was performed
with the latest version of the Monte Carlo RT code by 
Whitney et al. (2003). 
Corrections made by adiabatic cooling/heating and advection were included, and so were the gas opacities.

The model was compared to the massive protostar G35.2-0.74N. 
At a distance of 2.2 kpc, radio continuum emission indicates there is a bipolar
outflow from this source. In recent SOFIA-FORCAST observations of the massive protostar G35.2-0.74N at 
31 and 37μm (Zhang et al. 2012, in prep.) both the near- and far-facing sides of the outflow can be seen. 
The latter is missing in mid-IR continuum. 
By fitting both the observed SED and the outflow-axis intensity profiles with the RT model, the bolometric 
luminosity was inferred to be $\sim 10^{5}$\,\lsun\ after correcting for foreground
extinction and the dependence of luminosity on inclination (flashlight effect). 

The fitting model suggests a massive $\sim$ 30\,\msun\ protostar is
forming from a $\sim$240\,\msun\ core with a high surface density of 
$\sim$1g/cm$^2$, via relatively 
ordered collapse and accretion, and driving powerful bipolar
outflows. These results seem to support the core accretion theory which predicts massive stars 
may form in similar ways to their low-mass counterparts. 
Simply extending the model to protostars with higher masses up to 100\,\msun, Zhang et al. 
noticed a shift of the SED peaks to shorter wavelengths and a change of mid-IR slope. 
The flashlight effect turns out to be huge for such sources. 
In the mid-IR, the luminosity from face-on view can
be higher than that from edge-on view by as much as 3 dex -- suggesting 
that a large correction 
factor needs to be applied when using IR observations for deducing 
source luminosities.

\section{Properties of VMS}
\label{sec:props}

Once a star has formed it starts to burn hydrogen on the ZAMS.  
For massive stars, the main sequence band probably consists of two
spectroscopically distinct groups of objects: the O stars, and the 
WNh stars. 
The first group are thought to form the normal massive star sequence with masses 
of up to at least $\sim$60\msun. For the most massive ``O stars on steroids''  
WNh (H-rich WR stars), their masses may be up to 300\,\msun. or higher. 

The transition mass between O and WNh 
is not exactly known but is probably lies somewhere 
in the range 60-120$\msun$. Of course these numbers are model (mass loss and $Z$) dependent. 
The WNh stars are presumably 
in close proximity to the Eddington limit, and so are their descendant LBV and 
classical WN, WC, WO Wolf-Rayet stars. Some of these objects are thought to be in an evolutionary 
phase just prior to their final demise.

At the end of this section, we also consider extra-galactic VMS properties, both
at very low (e.g. IC 1613) and high metal content (e.g. M33).

\subsection{WNh H-rich Wolf-Rayet stars -- ``O stars on steroids''}

WNh stars were discussed by Crowther and Hamann. In their analysis they make 
use of non-LTE model atmospheres, such as CMFGEN (Hillier \& Miller 1998) 
and PoWR (Hamann et al. 2006), whilst FASTWIND  (Puls et al. 2005)
is often used for normal O stars with weaker winds. 
These are 1D spherical symmetric codes that 
include line-blanketing and stellar winds allowing for micro-clumping (optically thick 
macro-clumping allowing for porosity is non-standard). Another important aspect of the analysis is that 
of accurate infrared photometry (see the infrared SpS on atmosphere modelling). 
This is especially true for those objects in the cores of dense clusters
such as R136. The upshot of recent analyses is that WNh stars are more luminous, and more 
massive, than previously thought (see Sect.\,\ref{sec:exist}). 
Hamann presented some early results from their spectral analysis of WN stars in the Magellanic Clouds 
(Hainich et al. in prep.) -- complementing the published Galactic study of 
Hamann et al. (2006). Whilst the LMC objects are found to be 
bright with log$L/\Lsun$ values in the range 5.3-5.9, Hamann warns that many of the 
brighter allegedly single WNh stars may actually be binaries, and not accounting for this fact may overestimate the 
luminosities of WNh stars.
  
\subsection{LBVs: unstable massive stars close to the Eddington limit}

Very massive stars are thought to evolve through the unstable Luminous Blue Variable phase, when enormous amounts of mass are lost. While LBVs
have been classically thought to be rapidly evolving massive stars in the transitory phase from O-type to Wolf-Rayet stars, recent studies have
suggested that LBVs might surprisingly explode prematurely 
as a core-collapse supernova (Kotak \& Vink 2006, Gal-Yam et al. 2007, Mauerhan et al. 2012). 
Such a striking result highlights that the evolution of VMS through the LBV phase is far from being understood.

Groh discussed the recent advances in understanding LBVs, in particular how
to distinguish them from the normal B supergiants and hypergiants (as discussed by Negueruela).
LBVs can be recognized either from Giant Eruptions like Eta Car and P Cygni, or through their S-Dor 
variability, sometimes imprinted on peculiar looking double-peaked absorption profiles, in just a single epoch spectrum (Groh \& Vink 2011). 
LBVs do not always present both S\,Dor and giant outburst phenomena, leaving room for 
quite a heterogeneous class of objects. 

Groh emphasized that LBVs are characterized on a phenomenological
basis and, therefore, LBVs are neither a spectroscopic nor 
an evolutionary classification. Particular emphasis was given to describe the main
properties of the S-Dor type variability (also performed by Stringfellow). 
These are changes in the hydrostatic radius and bolometric luminosity. 
Finally, the role of rapid rotation on LBVs was discussed (Groh et al. 2009).

\subsection{The Galactic WC stars}

WC star spectra are well known for their broad emission lines from
helium, carbon and oxygen. Due to the absence of hydrogen these stars
have to be core-helium burning and have as such been identified as 
the late evolutionary stage of massive and very massive stars. Sander et al. (2012) 
analyzed the optical and UV spectra of over 50 Galactic WC single stars and 
derived their parameters including their mass-loss rates.
Sander et al. showed that the positions of the Galactic WC stars
in the HR-diagram do not fit with the assumption that the most massive 
stars will pass the WC stage. Instead Sander et al. argue that their 
results indicate that WC stars come from an initial mass range between 20 and 
somewhere around 50\,\Msun. The stellar evolution models of Vanbeveren et al. (1998) that include
enhanced mass loss during the red sugergiant (RSG) phase appear to properly 
account for the location of the WC stars in the observed HR-diagram.
It also seems that stars with higher initial masses do not reach the WC
stage but instead explode after passing a WNL and probably an LBV stage.

\subsection{WO stars}

Sander et al. also showed that the Galactic WO2 stars WR 142 and WR 102 have significantly 
different parameters from the WC stars. The WO positions in the HR diagram suggest that these stars
could be close to or already in the stage of carbon burning which makes them
interesting SN Ic candidates (Georgy et al. 2012; Yoon et al. 2012).

Tramper et al. presented VLT/X-Shooter spectroscopy of DR1, a WO3 star in 
the low-metallicity galaxy IC 1613. A preliminary spectroscopic analyis 
using CMFGEN indicates a high temperature
(of $\sim$150 kK) which is also supported by the very strong nebular He II emission. The oxygen
abundance does not seem to be enhanced compared to values found for early-type WC stars,
suggesting that the strong oxygen emission is likely a temperature effect, rather than being
caused by an increased oxygen abundance.

\subsection{Rapidly rotating WR stars}

The WR rotation issue is especially relevant in view of the suggested link
between rotating WR stars and long-duration GRBs. 
WR 2 (WN2) is a well known oddball. It is the most compact  
and the hottest ($\sim$ 140 kK) pop I WN star known in the Galaxy. It is also 
one of the best candidate for strange-mode
pulsations (Glatzel et al. 1999), but they are not observed in 
photometric observations. The spectrum displays bowler-hat shaped emission lines, in
contrast with the more normal Gaussian or even flat-top and triangular profiles of most other WR winds.
Remarkably, Hamann et al. (2006) have analyzed WR 2 using their latest 
model-atmosphere code, and the model spectrum 
failed to reproduce its line-profiles, unless it was folded with a rotation curve near the break-up limit, 
i.e. 1900 km/s. 

However, Chen\`{e} presented the polarized spectrum of WR 2, which 
shows no sign of wind asymmetry expected for such rapid rotation. 
Interestingly, WR 2 appears to display clumps that are moving in a 
similar fashion as in the optically thinner wind of WR 3 
(Chen\'{e} et al. 2008). 
Hence, the shape of WR 2 spectra line cannot be the result of extreme 
opacity either.\\

\subsection{Massive stars in the low metallicity galaxy IC1613}

Because low metallicity environments may favour the formation of massive and very massive stars, we 
switch our attention to extra-galactic properties. 
Herrero discussed the selection of massive star candidate stars in the low-Z galaxy IC\,1613 from the catalog of 
Garcia et al (2009). IC\,1613 has a metallicity Z= 0.13 Z$_\odot$ 
according to the analysis of B-supergiants by Bresolin et al. (2007), or Z= 0.08$-$0.15 
Z$_\odot$ according to several analyses of HII regions (see Herrero et al. 2012).
Herrero et al. observed the selected stars with OSIRIS/GTC at a resolution $R\,=\,1000$ 
to determine their spectral types, as a previous step for a more detailed analysis.
They were able to classify 12 new OB stars and confirmed one more known O-type star. 

The spectra of the O stars were good enough for quantitative analysis (albeit with
errors slightly larger than in typical analyses, see Repolust et al. 2004). 
Combining these results to those from the literature (Tramper et al. 2011; Herrero et al. 2012), 
Herrero presented the first effective temperature scale for sub-SMC metallicities. 
This temperature scale is slightly hotter than that derived for SMC stars
from the data of Massey et al. (2009), Mokiem et al. (2007), and Trundle et al. (2007).

\subsection{Stellar abundances from massive stars in M33}

With its distance less than a Mpc (Bonanos et al. 2006) and its favourable 
inclination angle, this makes M33 an ideal galaxy to study the 
chemical evolution in spiral galaxies. Not long ago, the only way to carry out detailed 
chemical analyses of nearby galaxies was through the quantitative studies of H II regions. 
However, the abundances derived from massive OB stars are the tracers of the present-day 
chemical composition, providing information that cannot be obtained H II regions (e.g. the silicon abundance). 
Moreover, a simultaneous characterization of the stellar parameters may address important aspects of 
their evolution, and in particular the role of environmental factors. 

Castro presented the results of a spectroscopic survey in M\,33 involving 
59 supergiants with spectral types between B9 and O9, and a quantitative analysis 
according to the steps described by Castro et al. (2012). A thorough 
comparison between optical spectra and 
new {\sc fastwind} (Puls et al. 2005) 
grids resulted in both stellar parameters and chemical composition. 
The parameters derived in conjunction with the evolutionary tracks of 
Brott et al. (2011) 
hints at the presence of evolved stars with masses in the range 
15 and 50 M$\odot$. New routines for deriving 
the chemical abundances automatically through a process optimization showed an oxygen distribution along M~33 that is 
compatible with previous H II region studies (e.g. Rosolowsky \& Simon 2008). 

\section{Mass loss mechanisms for VMS}
\label{sec:massloss}

The evolution of Very Massive Stars 
is presumably dominated by mass loss, which
thus needs to be understood both qualitatively and quantitatively. 
Joachim Puls reviewed 
different mass-loss mechanisms relevant in this context. Because of
the high luminosities, only radiation-driven mass loss was considered, and 
time-dependence, rotation and magnetic fields were not accounted for.\\

\subsection{Theoretical considerations}

\paragraph{\bf Basic considerations.} The equation of motion for the
transonic/supersonic regime of an expanding wind can be approximated
by 
\begin{displaymath}
v\Bigl(1-\frac{a^2}{v^2}\Bigr)\frac{{\rm d}v}{{\rm d}r} \approx
g_{\rm grav}(r) + g_{\rm rad}^{\rm tot}(r) =-\frac{GM}{r^2}(1-\Gamma(r)),
\quad \Gamma(r) = \frac{\bar \kappa(r) L_\ast}{4\pi GMc} = \frac{\bar
\kappa(r)}{\sigma_{\rm e}}\Gammae
\end{displaymath}
where all quantities have their usual meaning, $a$ is the isothermal
sound speed, $g_{\rm rad}^{\rm tot}$ is the {\it total} radiative
acceleration (lines + continuum), $\bar \kappa$ is the flux-mean opacity per unit mass, and
$\Gamma$ the corresponding Eddington parameter (\Gammae\ w.r.t.
electron scattering only). At the sonic point, $r_s$, $v=a$, and thus 
$g_{\rm rad}^{\rm tot} = -g_{\rm grav}$ implying $\Gamma(r_s)=1$.
To allow for an accelerating wind, $\frac{{\rm d}v}{{\rm d}r}|_s>0$,
which then requires $\frac{{\rm d}\bar \kappa}{{\rm d}r}|_s>0$ and
$\Gamma(r) < 1$ below and $\Gamma(r) > 1$ above the sonic point,
respectively.\\

\noindent
\paragraph{\bf Photon tiring limit.}
The mechanical luminosity of the wind at `infinity' is given by 
\begin{displaymath}
L_{\rm wind}=\mdot\Bigl(\frac{\vinf^2}{2}+\frac{GM}{R}\Bigr)=
\mdot\Bigl(\frac{\vinf^2}{2}+\frac{\vesc^2}{2}\Bigr) \quad \mbox{with}
\quad \vesc=\sqrt{\frac{2GM}{R}},
\end{displaymath}
and the maximum mass-loss rate follows from the condition that $L_{\rm
wind} = L_\ast$ (when the star would become invisible): $\mdot_{\rm max} =
2L_\ast/(\vinf^2+\vesc^2)$. Following \citet{OwockiGayley97}, \mtir\
then is the maximum mass-loss rate when the wind just escapes the
gravitational potential, with $\vinf \rightarrow 0$, and is much
larger than typical mass-loss rates from line-driven winds,
\begin{displaymath}
\mtir=\frac{2L_\ast}{\vesc^2}=0.032 \frac{\msun}{{\rm yr}}
\frac{L_\ast}{10^6\lsun} \frac{R}{\rsun} \frac{\msun}{M} = 
0.0012\frac{\msun}{{\rm yr}} \Gammae \frac{R}{\rsun}.
\end{displaymath}\\

\noindent
\paragraph{\bf Continuum driven winds.} To drive a wind by pure
continuum acceleration ($\Gamma = \Gamma^{\rm cont}$) when the
photosphere is sub-Eddington ($\Gamma(r) < 1$ for $r<r_s$) requires
substantial fine-tuning to reach and maintain $\Gamma^{\rm cont}(r)
\ge 1$ for $r \ge r_s$, and is rather unlikely. Anyhow, in such a
situation $g_{\rm rad}^{\rm cont}$ is almost density-independent, and
large mass-loss rates could be accelerated, only limited by photon
tiring, which needs to be considered in the equation of motion (for
details, see \citealt{OwockiGayley97}).\\
%\begin{displaymath}
%\frac{r^2}{GM}v\Bigl(1-\frac{a^2}{v^2}\Bigr)\frac{{\rm d}v}{{\rm d}r} \approx
%-1+\Gamma^{\rm
%cont}(r)\Bigl[1-m\bigl(\frac{v^2}{\vesc^2}+1-\frac{R}{r}\bigl)\Bigl],
%\quad m=\frac{\mdot}{\mtir}
%\end{displaymath}

\noindent
\paragraph{\bf Super-Eddington winds.} If, on the other hand, the
complete atmosphere is super-Eddington, $\Gamma(r) >1$, continuum
driving (mostly due to electron-scattering) might become possible.
When atmospheres approach or exceed the Eddington limit, non-radial
instabilities arise making them inhomogeneous (clumpy). Photons on
their way out avoid regions of enhanced density, and the medium
becomes {\it porous}. In this case, the photospheric radiative
acceleration decreases compared to an unclumped medium, leading to an
{\it effective} Eddington parameter {\it below unity}. In the outer
regions, where the clumps become optically thin due to expansion,
porosity decreases and $\Gamma_{\rm eff}^{\rm cont} \rightarrow
\Gamma(r) >1$. Thus, an accelerating wind can be initiated
\citep{Shaviv00, Shaviv01a, Shaviv01b}. \citet{Owockietal04}
expressed the effective opacity in terms of a `porosity length',
and showed that associated mass-loss rates can become substantial when
this length is on the order of the pressure scale height, amounting to
a few percent of the tiring limit. Invoking a power-law distributed
porosity length, they showed that the `observed' mass loss from the
giant outburst of $\eta$~Car might be explained by this
(metallicity-independent!) mechanism, and that \mdot\ scales with 
%
%\begin{displaymath}
%\Gammaeff(r)=\frac{\kappa_{\rm eff}(r)}{\bar \kappa(r)}\Gamma, \mbox{
%with} \bar \kappa \ge \sigma_{\rm e} \approx 0.34 \frac{\rm g}{\rm cm^2}
%\end{displaymath}
%
%\begin{displaymath}
%\frac{\kappa_{\rm eff}(r)}{\bar \kappa(r)}=\frac{\rho_{\rm
%c}}{\rhob}\bigl(1-{\rm e}^{-\rhob/\rho_{\rm c}}\bigr) \mbox{ with} 
%\rho_{\rm c} = \frac{1}{h\bar\kappa}
%\end{displaymath}
%
%
%condition at sonic point:
%\begin{displaymath}
%\Gammaeff(r)=\frac{\kappa_{\rm eff}(r)}{\bar \kappa(r)}\Gamma = 1
%\rightarrow
%\frac{\rho_{\rm c}}{\rhob}\bigl(1-{\rm e}^{-\rhob/\rho_{\rm
%c}}\bigr)|_{r_s} = \frac{1}{\Gamma(r_s)}
%\end{displaymath}
%
%\begin{displaymath}
%\mdot=\Bigl(1-\frac{1}{\Gamma^2}\bigr)\frac{H}{h}\frac{L_\ast}{ac}
%\mbox{ and } \frac{\mdot}{\mtir}\approx \frac{\vesc^2}{2ac}=
%\frac{0.032}{a_20}\frac{M}{\msun}{\rsun}{R}
%\end{displaymath}
%\begin{displaymath}
$\mdot \propto \Gamma^{1/\alpha_p-1}$,
%\end{displaymath}
as long as the exponent of the power-law, $\alpha_p <1$ and $\Gamma >
3{\ldots} 4$.\\

\begin{figure}
 \includegraphics[width=8cm, angle=90]{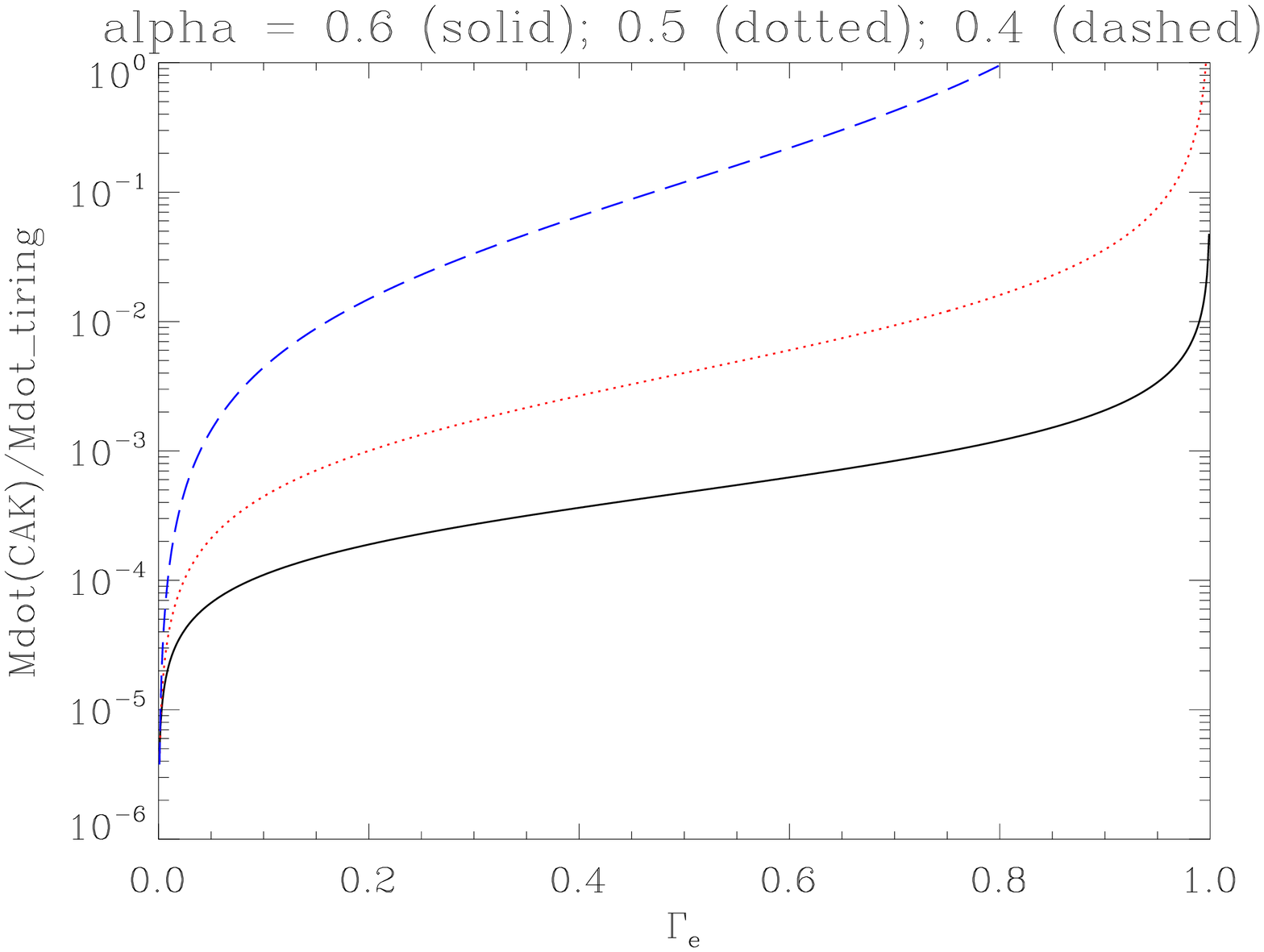}
  \caption{\mdot/\mtir\ for a typical O-star wind with \vesc =
  600~kms$^{-1}$, as a function of \Gammae\ and $\alpha$.}
\label{mdotcak}
\end{figure}

\noindent
\paragraph{\bf Line-driven winds.} The standard theory of
line-driven winds (\citealt{CAK} and later refinements) assumes
that the continuum is still optically thin at the sonic point (valid
for OB-stars, A-supergiants and LBVs in their quiet phase), and that
$\Gamma^{\rm cont} < 1$ everywhere, with $\Gamma^{\rm cont}
\rightarrow \Gammae$ for $r \ge r_s$. The radiative acceleration
exerted on a shell of mass $\Delta m = 4\pi r^2 \rho$ by {\it one}
optically thick line can be expressed as 
\begin{displaymath}
g_{\rm rad}^{\rm one\, line} = \frac{\Delta P}{\Delta t \Delta m} \propto
\frac{L_\nu \nu}{4\pi r^2}\frac{{\rm d} v}{\rho {\rm d} r},
\end{displaymath}
where $\Delta P$ is the transferred momentum, and the term ${\rm d}v$
arises because of the Doppler-shift within ${\rm d}r$. Summing up over
all lines using a line-strength distribution function and accounting for
optical depth effects, the total line acceleration results in
\begin{displaymath}
g_{\rm rad}^{\rm all\,\, lines} \propto N_0 \frac{L_\ast}{4\pi r^2}
\Bigl(\frac{{\rm d} v}{\rho {\rm d} r}\Bigr)^\alpha \rightarrow
N_0 \frac{L_\ast}{4\pi r^2} \Bigl(\frac{4 \pi}{\mdot}\Bigr)^\alpha
\Bigl( r^2 v \frac{{\rm d} v}{{\rm d} r}\Bigr)^\alpha,
\end{displaymath}
with $N_0$ the effective number of driving lines (depending on
spectral type and metallicity), and $0 < \alpha < 1$ related to the
slope of the line distribution function. Inserting this expression into the
equation of motion and neglecting photon tiring, a unique (maximum) mass-loss rate can be calculated,
\begin{displaymath}
\mdot \propto N_0^{1/\alpha}
L_\ast^{1/\alpha}\bigl(M(1-\Gammae)\bigr)^{1-1/\alpha} =
N_0^{1/\alpha} L_\ast
\Bigl(\frac{\Gammae}{1-\Gammae}\Bigr)^{1/\alpha-1} =
\Bigl(\frac{N_0 \Gammae}{1-\Gammae}\Bigr)^{1/\alpha}M(1-\Gammae),
\end{displaymath}
which dramatically increases for $\Gammae \rightarrow 1$ (in this
case, photon-tiring needs to be accounted for).
On the other hand, the terminal velocity scales with
\begin{displaymath}
\vinf \propto \vesc^{\rm eff} = \sqrt{\frac{2GM(1-\Gammae)}{R}}
\rightarrow 0 \mbox{ for } \Gammae \rightarrow 1.
\end{displaymath}
Figure \ref{mdotcak} displays the ratio of $\mdot/\mtir$ as a function
of $\Gammae$, and shows that this ratio is strongly sensitive to
$\alpha$. For $\alpha = 0.4$ (somewhat lower than the typical value
for OB-stars), the tiring limit would already be reached at \Gammae\ =
0.8.\\
%\begin{displaymath}
%\mdot \vinf (R/\rsun)^{1/2} \propto N_0^{1/\alpha'} L_\ast^{1/\alpha'} 
%\end{displaymath}

\noindent
\paragraph{\bf Optically thick winds.} The large mass-loss rates from
WR-stars, being typically a factor of 10 higher compared to OB-star
winds at the same luminosity, cannot be explained by the standard
theory from above. The observed terminal velocities (similar to
OB-stars) can be reached only when line-overlap effects become
efficient. In such dense winds, the ionization
equilibrium decreases outwards, and photons on their way out can
interact with lines from different ions, whilst any `gaps' between
lines (as present in OB-stars because of an almost frozen-in
ionization) are closed (see \citealt{LucyAbbott93, Springmann97}). The
initiation of the mass loss, on the other hand, is supposed to rely on
the condition that the winds are already optically thick at the sonic
point, and that the (quasi-static) photospheric line acceleration due
to the iron peak opacities around 150 kK (for WNEs) or 40 kK (for
WNLs) is sufficient to overcome gravity \citep{NugisLamers02}.
\citet{Graf05,Graf08} calculated
self-consistent models for WNEs and WNLs, and showed that this
mechanism actually allows for large \mdot, where the major prerequisite 
is a high $\Gamma$. Such optically thick winds might be
present also in VMS (\citealt{Vinketal11} and these proceedings), although 
\citet{Pauldrachetal12} argue that VMS winds might remain optically thin.
%
%
%\begin{displaymath}
%\frac{\bar \kappa}{\sigma_{\rm e}}\Gammae|_{r_s} = 1 \rightarrow \bar
%\kappa = \frac{\sigma_{\rm e}}{\Gammae}
%\end{displaymath}
%
%\begin{displaymath}
%\mdot \propto T_s^{4.5}\frac{R_s^3}{M} =
%\frac{T_s^{4.5}}{\Teff^4(R_s)}R_s\Gammae
%\end{displaymath}
%
%\begin{eqnarray}
%\log \mdot_{\rm WNL} &=& \beta(Z)\log(\Gammae-\Gamma_0(Z))-3.5\log T_\ast
%+0.42\log L_\ast -0.45 X_{\rm H} + \mbox{const} \\
%T_\ast &=&\Teff(\tau_{\rm Ross} = 20)=\Bigl(\frac{L_\ast}{\sigma_{\rm B}
%4\pi R_\ast^2}\Bigr)^{1/4};  \quad \beta(Z_\odot) \approx 2 \nonumber
%\end{eqnarray}
%

\subsection{Monte Carlo mass-loss rates}

Vink discussed three relatively new aspects concerning  
mass-loss calculations from the Monte Carlo 
method (Abbott \& Lucy 1985) - as 
previously used to predict $\dot{M}$ for canonical 
OB-type stars (e.g. Vink et al. 2000).
The first one concerned the wind dynamics. 
Until 2008, the methodology was semi-empirical, as 
a velocity law was assumed that reached 
a certain empirical $v_{\infty}$. 
M\"uller \& Vink (2008) suggested a line-force 
parametrization that explicitly depends on radius 
(rather than the velocity gradient, as in CAK theory), and 
predicted $v_{\infty}$ values in reasonable agreement with observations.
Muijres et al. (2012) tested the M\"uller \& Vink 
approach, and as both methods gave similar results, 
it was used in the following.

Secondly, a new parameter space was probed, i.e. that of the VMS. 
Vink et al. (2011) $\dot{M}$ 
predictions show a {\it kink} in the $\dot{M}$ - $\Gamma$ relation. 
For ``low'' $\Gamma$ 
optically-thin O-star winds, the $\dot{M}$ $\propto$ 
$\Gamma^{x}$ relation is shallow, with $x$ $\simeq$2, whilst there 
is a steepening at high $\Gamma$, with
$x$ $\simeq$5. At high $\Gamma$ the objects show optically thick 
WR-like winds, with optical depths and wind efficiencies above unity. 
Gr\"afener et al. (2011) provided empirical evidence for 
such a steep exponent ($x$ $\simeq$5), but 
there are still issues with the predictions of 
absolute $v_{\infty}$ values 
in this high $\Gamma$ range. Critical comparisons between observations 
and theory
are underway by Bestenlehner et al. 
in the context  of the VFTS survey (Evans et al. 2011)

In another study Vink \& Gr\"afener (2012) 
calibrated wind mass-loss rates using an 
analytic method to find that the wind efficiency number 
equals unity right at the transition point between 
optically thin and thick winds: $\eta = \tau = 1$. 
Application of this relation to the most massive 
stars in the Arches cluster suggests 
there is little room for  
additional mass loss during e.g. LBV eruptions, and current wisdom
would   
suggest that PSN explosions are unlikely, unless one 
were to move to lower $Z$ galaxies (e.g. Langer et al. 2007, 
Yoshida \& Umeda 2011, Yusof et al. in prep.).\\

\subsection{Alternative mass loss: eruptions and mass transfer}

Solar metallicity VMS likely “evaporate” as the result of stellar wind mass loss. 
However, alternative mass loss may also be important, especially 
for the lower initial mass and sub-solar metallicity ranges. 
Furthermore, we know Eta Car analogs and supernova impostors exist in 
external galaxies (e.g. Van Dyck et al. 2005, Pastorello et al.
2010, Kochanek et al. 2012), but quantitative estimates on the integrated amount of
such eruptive mass loss are hard to obtain as both the eruption frequency, and the amounts of
mass lost per eruption span a wide range with LBV nebular mass estimates
varying from $\sim$0.1\,\msun\ in P\,Cygni 
to $\sim$10\,\Msun\ in Eta Car (Smith \& Owocki 2006). 

The energies required to produce such giant mass eruptions are very high
($\sim10^{50}$ erg), and their energy source is unknown. 
Soker (2004) discussed that the energy and
angular momentum required for Eta Car's great eruption cannot be 
explained with a single-star scenario. \\

There is a growing amount of evidence that the most massive stars are oftentimes
found in binary systems, and binary evolution with mass loss is 
pursued by many groups around the globe (Vanbeveren 1998; Eldridge et al. 2008). 
What has become particularly clear from 
recent spectroscopic radial velocity surveys (e.g. Sana et al. 2012) is that 
there is a particularly large number of short-period binaries, which 
might merge still during core 
H burning, and subsequently evolve as seemingly single stars. For these 
reasons one of the most famous recent phrases in the massive star community has been
``binary stars might actually be the best single stars'' (de Mink et al. 2011).
However, also after the main sequence, there are still many physical processes 
involving mass loss through Roche lobe overflow (see e.g. Langer 2012) and 
common envelope evolution (Ivanova et al. 2012), which remain as yet ill-understood.\\

\subsection{Mass-loss diagnostics}

The traditional ways of determining mass-loss rates of (very) massive 
stars involve recombination lines (such as H$\alpha$ and He {\sc ii} 4686), as 
well as radio and sub-mm continuum measurements that measure the amount of 
free-free emission (Wright \& Barlow 1975, Lamers
\& Cassinelli 1999). Especially the free-free method may be used in the near future
with new facilities such as ALMA coming online. The drawback of the above 
diagnostics is that they depend on an uncertain amount of wind clumping (Puls et al. 2008).
The unsaturated resonance lines of trace elements, such as P\,{\sc v}, 
located in the far UV, has been considered a more accurate tool for mass-loss
diagnostics (e.g. Fullerton et al. 2006), because their formation depends linearly
on density such that inhomogeneities average out. However, it was 
shown by Oskinova et al. (2007) that the formation of resonance lines is also 
affected by wind clumping, if the line opacity makes the individual clumps
optically thick. Neglecting this effect may lead to {\it under}estimations of 
the true $\dot{M}$ (see also Sundqvist et al. 2010).

For these reasons it is important to (i) gain a greater understanding 
of both the physics and the diagnostics of wind clumping throughout the stellar atmosphere 
and wind, ideally as a function of radial distance, and 
(ii) to develop diagnostics that are {\it not} dependent on wind clumping.

\subsubsection{X-ray diagnostics for VMS}

Massive stars of most (but not all) spectral types are sources
of X-ray emission. In single stars, the X-rays most likely originate in
the gas heated by the strong shocks resulting from the line-driven
instability of stellar winds (e.g. Lucy 1980, Owocki et al. 1988). 
Therefore,  the properties of X-ray
emission are sensitive to the wind driving mechanism. Because of
the proximity of VMSs to the Eddington limit, the details of wind driving and
line-driving instability growth may be different from lower mass massive stars. 

The X-ray luminosity of VMS stars is challenging to predict. The X-ray
luminosity of Galactic OB stars follows the trend $L_{\rm X}\propto
10^{-7}L_{\rm bol}$. While some binary O stars with colliding winds
have X-ray luminosity significantly higher than $10^{-7}L_{\rm bol}$,
the majority of O star binaries follow this correlation 
as well (Oskinova 2005, Naze 2009). In some cases,
the binary O stars have X-ray luminosity significantly lower than 
the expected for a single star of similar spectral type. Oskinova produced 
a diagram showing the dependence of the X-ray luminosity in binary O stars on the period. 
No correlation was seen, and the short period binaries can have
low X-ray luminosity. 

Oskinova concluded that a low X-ray luminosity cannot
serve as a robust argument against a binary nature of an O star. In other words, a 
binary luminous VMS could potentially have a low $L_{\rm X}$. However, the story 
for the WR-like VMS binaries (with strong winds) might be different 
from the weaker-winded O-star binaries.\\

\subsubsection{A new wind measurement approach using X-rays from 
colliding wind binaries}

Sugagawa presented {\it Suzaku} observations of the WR binary WR 140, 
taken at four different times around periastron passage in 2009 January. The X-ray
spectra changed in shape and flux at each phase. As periastron approached, the column
density of the low-energy absorption increased, indicating that the emission from
the wind-wind collision plasma was absorbed by the dense WR wind. The luminosity of
the dominant hot component from the wind-wind collision is not inversely proportional
to the (variable) distance between the two stars. In the case of the mass-loss ratio
$\dot M_{\rm O}/\dot M_{\rm WR}=$0.04, Sugagawa could explain this discrepancy if the O-star
wind collides with the WR wind before it has reached its terminal velocity, leading
to a reduction in its wind momentum flux.  
Sugagawa presented these mass-loss rates, which were
calculated using the absorptions and variations of the spectra (Sugawara et al. 2012, submitted). 

Daminelli showed that the He II 4686 line in Eta Carinae displays two peaks before periastron in good
correlation with the X-ray intensity. There even is a third peak, in close coincidence with
periastron, which is anti-correlated with X-rays intensity. This may be interpreted as a
collapse in the wind-wind collision structure, when most of the
energy escapes in the extreme UV, which would be possible if the eccentricity is larger than e$>0.9$.

What is clear is that the approach presented by Sugagawa using X-ray observations is applicable to 
other massive CWBs with elliptical orbits. In addition, unexpected X-ray brightening of 
very massive CWBs (such as WR 21a) may be helpful for understanding VMS mass loss.\\

\subsubsection{Mass loss at very low metallicity}

Whilst there are still several uncertainties in our empirical knowledge of 
Solar metallicity mass loss, rates at low $Z$ may provide additional constraints 
on the driving mechanisms. 
In this context, Tramper 
presented the results of a quantitative spectroscopic analysis of VLT/X-Shooter
observations of six O-type stars in the low-metallicity galaxies IC 1613, WLM and
NGC 3109 (Tramper et al., 2011; Herrero et al. 2011, but see also 
Herrero et al. 2012). The obtained stellar and wind parameters can be used to
probe the mass loss versus metallicity dependence at metallicities below that of the SMC. 
Tramper et al. compared their derived mass-loss rates with the empirical results from Mokiem et al. (2007) for the
Galaxy, LMC and SMC, and with the theoretical prediction from Vink et al. (2001), and argued 
that the mass-loss rates appear to be higher than expected.

It is clear that the analysis of a larger sample of stars at sub-SMC metallicities 
is needed to confirm or disprove these results.\\

\section{VMS Evolution and Fate}
\label{sec:evol}

Alexander Heger (with Woosley \& Chen) 
presented the fourth review talk on (very) massive star evolution. 
An introduction to massive star evolution 
can be found in \citet{2002RvMP...74.1015W, 2012ASSL..384..299H}.
One of the most exciting questions concerning massive stars is how they will die,
which of them will explode, and how. A range of outcomes is possible 
in terms of observational signatures, but the key question is 
how massive the star is at the time of death.
Therefore, according to Heger understanding the mass-loss rates of 
massive stars is one of the key uncertainties that needs to be resolved.  

Other stumbling blocks along the way involve rotation and binary 
evolution. For the sake of simplicity, Heger only discussed single stars. 
However, it is clear that some close binaries will merge early during their 
evolution, already during central hydrogen burning. These would 
likely evolve in a fashion similar to that of single
stars - though likely rapidly rotating.  Rotation however is a 
second parameter that makes the picture more
complex. So for the sake of simplicity Heger dealt ``just'' with 
single stars and deferred the reader to the rotation works
of \citet{2006A&A...460..199Y, 2012A&A...542A.113Y}. 
In some aspects, rapidly rotating
single stars may evolve in a fashion similar to non-rotating stars of
a different (usually higher) mass, maybe if mass-loss rates are
varied (artificially) similar results can be obtained.  

For the final death, the explosion mechanism and magnitude, knowing the
size of the (helium) core is the key ingredient. 
Heger discussed outcomes in terms of initial masses for Pop III stars where it 
was also assumed there is no mass loss by stellar winds, taking the key 
uncertainty out of the equation, and hence allows one to draw a clearer
picture -- though obviously not realistic. However, it allows one to
understand and formulate the possible outcomes.

Supernovae from lower mass massive stars generally produce neutron stars
within the 'classical' core collapse SN scenario. For
higher masses, again really meaning for higher mass cores at the end
of evolution, one eventually has to deal with core structures that no longer
allow for an efficient explosion and expect the star to
collapse to a black hole. Depending on the rotation of the star, a
GRB may result. Again, this may require binary stars or rapidly
rotating single stars, but here the focus is on the dependence on
star/core mass for \emph{possible} outcomes.  A recent review on such
possible outcomes can also be found in \citet{2012ApJ...752...32W}.  

When the mass of the star exceeds some $90-100\,\Msun$
($\sim42\,\Msun$ He core), pulsational pair instability (Puls-PSN) can occur. 
These are violent nuclear-powered pulsations during the final 
burning stages, usually powered by oxygen or silicon
burning. They may produce as much energy as a SN, or more. However, 
it could also be significantly less. Heger discussed this in more detail (see below). 
After the pulsing is over and the star forms an iron core, the fate should be
similar to the more massive stars discussed above.
For higher initial masses, above some $140\,\Msun$ ($65\,\Msun$ He
core) the first pulse is already powerful enough to disrupt the star
and a pair instability (PSN) results; for even higher initial masses,
above some $260\,\Msun$ ($133\,\Msun$ He core) it is expected that the
star collapses to a black hole. A brief discussion on what
could happen in that case may be found in \citet{2001ApJ...550..372F}.

One of the striking problems with the PSN theory is that
they produce a unique chemical signature - at least for the non-rotating
primordial stars.  Simulations of the formation of the first stars
suggest that they should have been rather
massive \citep[e.g.,][]{2002Sci...295...93A}, hence making PSN, and
their nucleosynthesis products should have been incorporated into the
next generation of stars, but to date, no stars with such 
nucleosynthesis pattern have been found.\\

\subsection{Pulsational pair instability (Puls-PSN)}

\begin{figure}
\includegraphics[width=\textwidth]{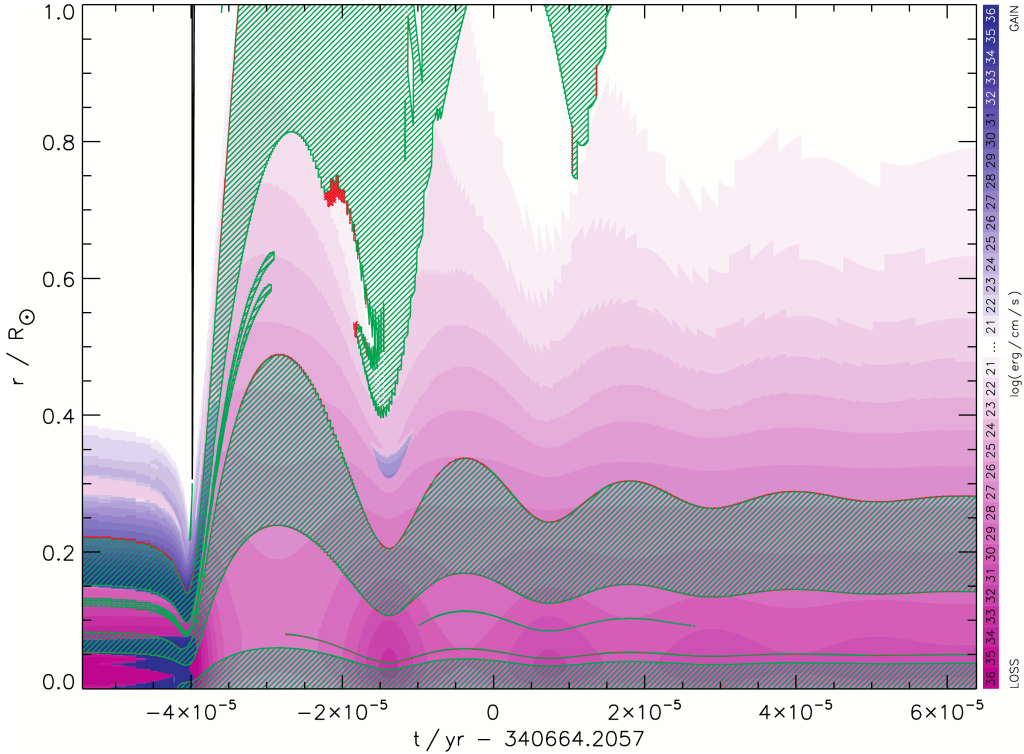}
  \caption{Energy loss (purple), energy generation (blue), and
    convection (green hatching) during a pulsational pair instability
    pulse of a $100\,\Msun$ star. 
The \textsl{x}-axis indicates the time, $t$, in years since the beginning of 
helium burning, the \textsl{y}-axis shows the radius
coordinate from the center of the star in solar radii.}
\label{fig:heger:PPSN}
\end{figure}

Figure~\ref{fig:heger:PPSN} depicts the dynamics of a Puls-PSN
simulation in 1D: after the pulse there is a ring-down phase in which
radiative dampening by neutrino losses brings the star back into
hydrostatic equilibrium. It is noted that the core of the star is
much more extended after the pulse than it was before (the reader may
follow convective boundaries as a reference).
In other words, the energy deposited by the burning during the pulse leads to an
expansion and cooling of the stars - as we know from textbooks, stars
have a negative (gravothermal) heat capacity. But in order to get
ready for the next pulse, it needs to get hot enough, and the energy needs to be lost
again. The more energetic, the cooler the core of the star will be.
But now the end stages of massive stars are dominated by neutrino
losses, and these are known to depend steeply with temperature. 
Therefore, the cooler the star, the longer it takes to cool.  

In extreme cases of rather powerful explosions, the temperature could 
drop enough such that the core has to cool on the classical 
Kelvin-Helmholtz time for radiative
losses from the surface, as it has become so cool that neutrinos
are no longer efficient. This can increase the recurrence time 
by several orders of magnitude. In summary, we generally expect that
more powerful pulses result in longer delay times, whereas weak
pulses result in short delay times for the next pulse. The
recurrence times can range from hours to more than 10,000 yr, and the
energies from $0.0001\,$B to several B (1 B = $1\times10^{51}\,$erg).

Why is all this relevant? 
Because we wish to understand whether some giant
eruption LBVs could be the result of Puls-PSN. 
In the case of Eta Car we dealt with a fairly weak explosion, 
as some of the H remained bounded to the star, so we would 
have expected the next pulse to happen soon after Eta Car's 
giant eruption. However, this was not observed, or so it appears.  
Therefore, at this point it seems unlikely that Eta Car's 
giant eruption was a Puls-PSN event.\\

\subsection{Stellar Envelope Inflation}

Alternative explanations for LBV outbursts and eruptions have been 
proposed over the years (see Humphreys \& Davidson 1994; Vink 2009).
Gr{\"a}fener et al. (2012) proposed the possibility that envelope inflation  
near the Eddington limit may play a key role in explaining the radius 
increases during S Dor cycles.
The peculiar structure of inflated envelopes, with an almost void region beneath a
dense shell could mean that many in reality compact stars are
hidden below inflated envelopes, displaying much lower effective
temperatures (see also Ishii et al. 1999; Petrovic et al. 2006). 

During the JD, Gr\"afener discussed the inflation 
effect for WR stars, whose observed radii are up to an order of magnitude larger than 
predicted by theory. 
Based on a new analytical formalism, he described the radial inflation as a function
of a dimensionless parameter $W$, which largely depends on the
topology of the Fe-opacity peak, i.e., on material properties. For
$W>1$, an instability limit is found for which the stellar envelope
becomes gravitationally unbound, i.e.\ there no longer exists a static
solution. Within this framework one may also be able to explain the
S\,Doradus-type instabilities for LBVs like AG\,Car (discussed by Groh during the meeting). 
Moreover, due to the additional effect of sub-photospheric clumping, it may be possible 
to bring the observed WR radii in agreement with theory (see Sander et al. results discussed earlier). 

It should be noted that stellar effective temperatures in the upper HR diagram may be 
strongly affected by the inflation effect.  This may have particularly
strong effects on the evolved massive LBV and WR stars just prior to
their final collapse, as the progenitors of supernovae (SNe)\,Ibc,
SNe\,II, and long-duration GRBs.\\

\subsection{3D Simulations of Thermonuclear Supernovae from VMS} 

Ke-Jung Chen (with Heger \& Woosley) 
presented results from numerical simulations of the demise of VMS 
with initial masses between 140\,\msun\ and 250\,\msun\ that can die as powerful 
PSN explosions. 
Chen et al. used CASTRO, a new 
multidimensional radiation-hydrodynamics code, to study the evolution of 
PSNe. The 3D simulations start with the collapse phase and follow the explosion 
until the shock breaks out from the stellar surface. Unlike the iron-core collapse 
SNe, PSNe are powered by thermonuclear runaway without leaving compact 
remnants. Much Ni is forged, up to 30\,\msun, and its decay energy powers the 
PSN luminosity for several months. During the explosion, the emergent fluid 
instabilities cause the mixing of PSN ejecta, and the amount of mixing is 
related to PSN progenitors. The red supergiant progenitors demonstrate 
strong mixing, altering the spectrum and light curves.\\

\section{Implications}
\label{sec:impl}

\subsection{Population synthesis models}

The implications for the existence of VMS may be far-reaching, as 
VMS may dominate both the kinetic wind energy input and the ionizing radiation 
in the Universe. Because of higher temperatures at lower metallicities, VMS may 
be increasingly UV bright. However, their higher luminosities might imply 
higher mass-loss rates and terminal wind velocities, which would 
account for an increased kinetic wind energy.\\

Voss presented his recent population models (Voss et al. 2009) 
which follow both the energy -- in the form of kinetic wind energy as well as radiation -- 
and the ejection of radio-active isotopes 
(such as $^{26}Al$) simultaneously.  
Below 120 $M_{\odot}$, stellar evolution calculations predict a
strong increase in the ejected mass of $^{26}$Al with stellar mass.
The ejection of $^{26}$Al from stars above this limit has
not been studied in detail, but as the mass-loss increases 
dramatically towards higher masses, it is reasonable 
to expect the $^{26}$Al to do the same. 
The short evolutionary timescale of VMSs mean that all
the $^{26}$Al is ejected 2-4 Myr after the star-forming episode, and
if present, the VMSs will dominate the signal for 2-5 Myr populations,
but due to the decay their signal will be negligible for older
populations. Comparing the $^{26}$Al signal from 2-5 Myr massive
open clusters to their 5-10 Myr counterparts is therefore a promising
way to probe the evolution of VMSs (Voss et al. 2010; 2012).\\

\subsection{Wolf-Rayet Stars in the Extraordinary Star Cluster NGC 3125-A1}

The powerful radiative and mechanical feedback from very massive stars ($\ge100\,$M$_\odot$) shape the evolution of star-forming galaxies and their
environments. 
Nearby galaxies ($\lesssim10$ Mpc) provide excellent laboratories for studying populations of such stars in sufficient detail in a variety of
astrophysical environments.

Aida Wofford (with Leitherer and Chandar) studied the massive star populations of clusters A1, A2, B1, and B2 in blue compact dwarf galaxy NGC 3125, which is located 11.5
Mpc away and has an LMC-like metallicity. It is unclear from past studies if cluster A1 hosts an extreme population of WR stars. 
In addition, the WR star populations of the other clusters are not well characterized. Wofford et al. obtained
HST/STIS 1200-9000 \AA~spectra of these four clusters, and higher resolution HST/COS 1200-1450 \AA~spectra of cluster A1, on which 
Wofford focused. 
The STIS spectrum of this cluster shows that the equivalent width of He\,II $\lambda$1640 is three times the mean of local starburst galaxies (Chandar et al.
2004) and three times the value of the strongest Lyman Break Galaxy (Erb et al. 2010). This suggests that A1 must have a large fraction of WR stars relative to
the number of O stars. Either A1 has a top heavy IMF or it contains a few massive stars with very strong winds. The COS spectrum of A1 shows the strongest O\,V
$+$ Fe\,V absorption feature at 1371 \AA\ from a starburst in the local universe. The O V line originates in the most massive stars and is sensitive to clumping in
the stellar wind. The analysis of the O\,V $+$ Fe\,V using CMFGEN stellar atmosphere models is underway.

\subsection{Nebular He\,II\,4686 emission: an indirect tracer 
of massive stars at low metallicities}

Shirazi (with Brinchmann) presented 
a carefully selected sample of 189 star-forming galaxies with
strong nebular He\,II\,4686 emissions in Sloan Digital Sky Survey Data (SDSS)
Release 7. They used this sample to investigate the origin of this high
ionization line in star-forming galaxies where the ionizing continuum almost 
certainly arises from massive stars. The current stellar population models can
predict He\,II\,4686 emission only for instantaneous bursts of 20\% solar
metallicity or higher, and only for ages of 4-5 Myr, the period when the
extreme-ultraviolet continuum is dominated by emission from 
WR stars.

Shirazi \& Brinchmann 
find however that 83 of the star-forming galaxies (70\% at
oxygen-abundance lower than 8.2) of their sample do not have WR features in
their spectra despite showing strong nebular He\,II\,4686 emission.
Nevertheless, at higher metallicities He\,II is always seen with WR
features. Shirazi went on to show that the stacked spectra of the non-WR He\,II emitters do
not show WR features either, which suggests that the 
non-detection of WR features in these galaxies is {\it not} due to low
signal-to-noise data, i.e. it is probably real.  

Shirazi proposed that a possible explanation for the
discrepancy between the model predictions and the observed data at very
low redshifts could be the result of a spatial offset between the 
location of the WR stars and the region where the He\,II emission arises from. 
Alternatively, as the non-WR He\,II emitters appear to be preferentially present in younger starbursts,
(quasi)-chemically homogeneous stellar evolution could provide 
a possible explanation, as this may lead to higher stellar 
temperatures, and perhaps result in an elevated He\,II emission even for 
main-sequence O stars. 

Shirazi \& Brinchmann 
are currently 
attempting to disentangle these explanations by analyzing higher 
signal-to-noise spectra
of a sub-sample of these galaxies that were followed up with the WHT.\\

\subsection{Very Massive Stars in I\,Zw\,18}

I\,Zw\,18 is a blue compact dwarf (BCD) galaxy with the lowest 
metallicity known (at 1/30 - 1/50 the solar value), and is therefore 
thought to be the best local galaxy template to galaxies at 
high redshift. 
Although I\,Zw\,18 is 15-19 Mpc away, Hubble/STIS imagery 
resolves stars in the galaxy. 

Heap showed that 
UV color-magnitudes diagrams 
indicate that the most massive stars in the northwest cluster 
of I\,Zw\,18 are as massive as 150\,\msun. 
Heap also showed that Hubble/COS 
far-UV spectra reveal that the mass-loss rates from stars in 
the NW cluster must be very low, as only the N\,V\,1240 doublet 
has a P\,Cygni profile, and it is very weak. The C\,IV\,1549 
doublet is resolved with an edge velocity of only 
$\sim$250 km/s. The emission component of 
the C\,IV doublet is quite possibly of nebular origin. Most of the 12 most 
luminous stars are bluer than ZAMS models, suggesting that the 
evolution of the most massive stars is affected by rapid 
rotation. A comparison of the observed CMD and FUV spectra 
with new evolutionary models including rotation will yield 
valuable information about chemical enrichment of the ISM, 
injected energy via ionizing radiation, and types of SN 
explosions in the NW cluster of I\,Zw\,18.\\ 

\section{Final words}

One of the key science goals for the James Webb Space Telescope (JWST) 
is going to be the identification of the first galaxies with Pop {\sc iii} stars. 
These objects may have been very massive (up to 1000\,\Msun). At this epoch 
black hole formation may have been more common than at the current time involving 
solar metallicities.
The first couple of stellar generations may also have been responsible 
for the reionization of the Universe: an important cosmological epoch that is 
soon to be probed via the 21\,cm line with instruments such as LOFAR.

With the E-ELT, {\it individual} (very) massive stars may be 
observed out to the Virgo cluster of galaxies in the Hubble flow, at ever larger distances, and 
for an increasing range of metallicities. 
Basic insights into mass loss at very low $Z$ involve issues such as the 
self-enrichment of metals through mixing (rotation, overshooting, magnetic fields, binarity) 
and mass loss in close proximity to the Eddington limit, which 
does not necessarily require metals, if the winds are continuum driven. 
We also need to consider the potential mass-loss-Z-dependence for the angular momentum evolution 
and the quest for pinpointing the progenitor stars of long-duration GRBs. 

The fate of massive stars is important for our understanding 
of the chemical enrichment of the Universe. 
Whilst the deaths of stars up to 15 \msun\ now seem to be pretty well 
known; those of VMS up to 300 \msun\ are as yet a complete mystery. 
Their fates may involve PSN, Puls-PSN, or just normal hydrogen-poor 
SNe Ibc, either with our without an accompanying GRB. 
Key unknown aspects in this quest involve the strength and 
geometry of the progenitor stellar wind. 
Equatorial winds would remove angular momentum during evolution, whilst  
spherical winds would not. Linear spectropolarimetry should become a particularly 
powerful tool to study the geometry of winds and disks during 
the evolution of (very) massive stars towards explosion. 

Similar arguments can be provided for the evolution of rotation rates 
during star formation. T Tauri and Herbig Ae/Be stars are 
the optically visible PMS up to $\sim$15 \Msun, and  
the formation of such objects is thought to proceed via 
disks. The latest observations and theoretical developments 
seem to suggest that ever more massive stars may form in similar fashions, i.e. 
via disk accretion, but before 
such far-reaching conclusions can be drawn, large stellar samples over the 
full IMF are needed. It is particularly important to explore the near IR part, 
in order to diminish the complications by dust extinction. 

GAIA is widely expected to provide key information about the massive star formation
history of our own Milky Way. 
This will involve issues of stellar dynamics, cluster 
formation, the role of massive star binarity, and the physical links 
between massive stars and their lower mass (T Tauri) siblings. 

Once we understand the upper IMF and the rotational/multiplicity 
properties of massive stars, we can concentrate on the evolution of the mass loss and 
rotation properties of VMS. Which objects could make PSNe and which ones might produce 
GRBs? May these transient phenomena become star-formation tracers at high redshift? 
If their luminosity functions could be mapped with respect to their redshifts, individual 
VMS may even allow astronomers to constrain galaxy formation models.

The future of VMS formation and evolution looks bright!

\begin{acknowledgments}
JSV would like to the the Royal Astronomical Society (RAS), the IAU, 
the UK Science and Technologies Facility Council (STFC) as well as the 
Northern Ireland Department of Culture Arts and Leisure (DCAL) for financial 
support. 
\end{acknowledgments}

\end{document}